\documentclass[aps,prd,preprintnumbers,superscriptaddress,nofootinbib,notitlepage,floatfix,preprint]{revtex4-2}
\usepackage[pdftex]{graphicx}
\usepackage{bm,latexsym,amsmath,amssymb,amsfonts,mathrsfs}
\usepackage{float}
\usepackage{color}
\usepackage{physics}
\usepackage{color}
\usepackage{comment}
\usepackage{here}
\usepackage{empheq}

\allowdisplaybreaks[1]
\usepackage[pdftex,colorlinks=true,linkcolor=blue,citecolor=cyan,backref=page]{hyperref}
\makeatletter
\let\MYcaption\@makecaption
\makeatother

\usepackage{subcaption}
\makeatletter
\let\@makecaption\MYcaption
\makeatother
\newcommand{\3}{\hspace{0.5pt}}
\newcommand{\1}{\hspace{1pt}}
\newcommand{\2}{\hspace{2pt}}

\renewcommand{\theequation}{\arabic{section}.\arabic{equation}}
\makeatletter
\@addtoreset{equation}{section}

\makeatother
\begin{document}
\title{New series expansion method for the periapsis shift}
%
\author{Akihito~Katsumata}
\email[Email: ]{a.katsumata@rikkyo.ac.jp}
\affiliation{Department of Physics, Rikkyo University, Toshima, Tokyo 171-8501, Japan}
\author{Tomohiro~Harada}
\email[Email: ]{harada@rikkyo.ac.jp}
\affiliation{Department of Physics, Rikkyo University, Toshima, Tokyo 171-8501, Japan}
\author{Kota~Ogasawara}
\email[Email: ]
{k\_oga@meiji.ac.jp}
\affiliation{Department of Physics, School of Science and Technology, Meiji University, Kanagawa 214-8571, Japan}
\author{Hayami~Iizuka}
\email[Email: ]{h.iizuka@rikkyo.ac.jp}
\affiliation{Department of Physics, Rikkyo University, Toshima, Tokyo 171-8501, Japan}
\date{\today}
%
\begin{abstract}
We propose a new series expansion method for the periapsis shift. The method formulates the periapsis shift in various spacetimes analytically without using special functions and provides simple and highly accurate approximate formulae. We derive new series representations for the periapsis shift in the Kerr and the Chazy-Curzon spacetimes by using the method, where the expansion parameter is defined as the eccentricity divided by the non-dimensional quantity that vanishes in the limit of the innermost stable circular orbit. That is to say, the expansion parameter denotes how eccentric the orbit is and how close it is to the innermost stable circular orbit. The smaller the eccentricity, the higher the accuracy of the formulae that are obtained by truncating the new series representations up to a finite number of terms. If the eccentricity is sufficiently small, the truncated new representations have higher accuracy than the post-Newtonian expansion formulae even in strong gravitational fields where the convergence of the post-Newtonian expansion formula is not guaranteed. On the other hand, even if the orbit is highly eccentric, the truncated new representations have comparable or higher accuracy than the post-Newtonian expansion formulae if the semi-major axis is sufficiently large. An exact formula for the periapsis shift of the quasi-circular orbit in the Chazy-Curzon spacetime is also obtained as a special case of the new series representation.
\end{abstract}
\preprint{RUP-24-14}
\maketitle
\newpage
\tableofcontents
\section{Introduction}\label{sec:intro}
General relativity, which has been the most successful theory of gravity so far, predicts some unique phenomena that are not found in Newtonian gravity. One such phenomenon is the periapsis shift. In Newtonian gravity, the trajectory of a test particle orbiting around an object is an elliptical orbit. On the other hand, in general relativity, the direction of the major axis of the orbit rotates by relativistic effect and this causes a shift of the periapsis, which is the point closest to the central object. This phenomenon is known as the periapsis shift. In particular, if the central object is the Sun, the shift is usually called the perihelion shift, and in a binary star system, it is often referred to as the periastron shift.

Recently, the supermassive compact object Sagittarius A* (Sgr A*) at the center of our galaxy has attracted the attention of many researchers. Although Sgr A* is often assumed to be a Kerr black hole (BH), the true identity of Sgr A* remains unclear. Motivated by this, there are a lot of studies that aim to discuss the true identity of Sgr A* by investigating alternative possibilities and the periapsis shifts around them. For example, periapsis shifts around naked singularities~\cite{Bini:2005dy,Bambhaniya:2021ybs,Ota:2021mub}, in dark matter distribution around a BH~\cite{Igata:2022rcm}, and in dark matter distribution with a dense core~\cite{Igata:2022nkt} have been investigated. Additionally, general formulae for the periapsis shift of a quasi-circular orbit in static spherically symmetric spacetimes are derived~\cite{Harada:2022uae}. Regarding recent observations, the periapsis shift of S2, which is a star orbiting around Sgr A*, has been observed by the Gravity Collaboration~\cite{GRAVITY:2020gka}. To discuss this and other observational results, it is important to understand the periapsis shift theoretically.

Since the successful explanation of the perihelion shift of Mercury by Einstein~\cite{Einstein:1916vd}, periapsis shifts in various spacetimes have been investigated. For example, the post-Newtonian (PN) expansion formulae for the periapsis shift in the Schwarzschild spacetime~\cite{Einstein:1916vd,Bini:2005dy, Vogt:2008zs,de2011estimating,poisson_will_2014,Tucker:2018rgy,He:2023joa}, in the Reissner-Nordstr\"{o}m spacetime~\cite{GongTian-Xi_2009,Heydari-Fard:2019pxd,Mandal2022}, in the Kerr spacetime~\cite{boyer_price_1965,Bini:2005dy,Vogt:2008zs,de2011estimating,He:2023joa}, and in the Kerr-Newman spacetime~\cite{C-Jiang2014,Heydari-Fard:2019kno} are well-known. However, the convergence of these formulae is not guaranteed in strong gravitational fields since they are derived under weak gravitational field approximations (i.e., the orbit is far from the central object). Alternative formulations of the periapsis shift to the PN expansion have been discussed in some papers. For example, a new formula for the periapsis shift for orbits with large eccentricities in the Schwarzschild spacetime has been deduced~\cite{Schmidt:2011xi}. In addition, methods in general stationary and axisymmetric spacetimes~\cite{He:2023joa} and in spherically symmetric spacetimes~\cite{Wang:2023cof} have been proposed. Recently, Walters has derived a new exact series representation for the periapsis shift in the Schwarzschild spacetime without using elliptic integrals~\cite{walters2018simple}, where the periapsis shift is analytically expressed as an infinite series and is also well approximated by a simple formula with high accuracy by truncating the series. In this paper, by generalizing Walters' work, we propose a new series expansion method for the periapsis shift, which formulates the periapsis shifts in various spacetimes analytically without using special functions and provides simple and highly accurate approximate formulae that are also valid in strong gravitational fields where the convergence of the PN expansion formula is not guaranteed. In the method, the expansion parameter is defined as the eccentricity divided by the non-dimensional quantity that vanishes in the limit of the innermost stable circular orbit (ISCO). That is to say, the parameter denotes how eccentric the orbit is and how close it is to the ISCO. This definition is analogous to the one in Walters' work. We consider the applications of the new method to the Kerr and the Chazy-Curzon spacetimes and derive new series representations for the periapsis shift. We also investigate the accuracies of the formulae that are obtained by truncating the new series representations up to a finite number of terms and compare them with those of the truncated PN expansion formula.

This paper is organized as follows. In Sec.~\ref{sec:peri_kerr}, we propose a new series expansion of the periapsis shift in the Kerr spacetime and derive a new series representation for the periapsis shift. We also investigate the dependence of the expansion parameter and the accuracy of the truncated new representation on the orbital parameters and compare them with that of the truncated PN expansion formula. In Sec.~\ref{sec:new_method}, we propose a general method for the new series expansion of the periapsis shift in various spacetimes by generalizing Walters' work. We reveal that the new series expansion of the periapsis shift in the Kerr spacetime which we propose in Sec.~\ref{sec:peri_kerr} is actually a specific application of the general method. As another specific example, we consider the application of the method to the Chazy-Curzon spacetime and derive a new series representation for the periapsis shift. The dependence of the expansion parameter and the accuracy of the truncated new representation on the orbital parameters are also investigated. In addition, we present an exact formula for the periapsis shift of the quasi-circular orbit in the Chazy-Curzon spacetime as a special case of the new series representation. Sec.~\ref{sec:sum} is devoted to the summary and conclusions. In Appendix~\ref{ap:Sch_kin}, the periapsis shift in the Schwarzschild spacetime is discussed. We briefly review Walters' work and another formulation by an elliptic integral. In Appendix~\ref{ap:kin_higher}, PN expansion formulae of the periapsis shift with higher order terms in the Schwarzschild, the Kerr, and the Chazy-Curzon spacetimes are presented. Throughout this paper, the geometrical units with $G=c=1$ and the sign convention of the metric $(-,+,+,+)$
are used unless otherwise stated.

\section{New series expansion of the periapsis shift in the Kerr spacetime}\label{sec:peri_kerr}
Recently, Walters has derived a new series representation for the periapsis shift in the Schwarzschild spacetime~\cite{walters2018simple}, where the periapsis shift is analytically expressed as an infinite series and is also well approximated by a simple formula with high accuracy by truncating the series. Further details are summarized in Appendix~\ref{ap:walters}. In this section, motivated by Walters' work, we propose a new series expansion of the periapsis shift in the Kerr spacetime. As we will see later, this expansion is actually a specific application of a general method we will propose in Sec.~\ref{sec:new_method}.

\subsection{New series representation for the periapsis shift} \label{subsec:deri_kerr}
The line element in the Kerr spacetime~\cite{Kerr:1963ud} in the Boyer-Lindquist coordinates~\cite{Boyer:1966qh} is given by
\begin{align}
  \dd{s}^2 &= -\biggl(1-\frac{2Mr}{\Sigma} \biggr) \dd{t}^2 - \frac{4 \1 aMr\sin^2 \theta}{\Sigma} \dd{t} \dd{\phi} + \frac{\Sigma}{\Delta} \dd{r}^2 \nonumber \\
  &\hspace{50truemm} + \Sigma \dd{\theta}^2 + \biggl(r^2+a^2+\frac{2Mra^2\sin^2 \theta}{\Sigma} \, \biggr) \sin^2 \theta \dd{\phi}^2 ,
  \label{met_kerr}
\end{align}
where $a := J/M$, $\Delta (r) := r^2 - 2Mr + a^2$, and $\Sigma (r,\theta) := r^2 + a^2 \cos^2 \theta$. The parameters $M$ and $J$ denote the mass and the angular momentum of the BH, respectively. We refer to $a$ as the Kerr parameter. Now we can assume $0 \leq a \leq M$ without loss of generality. Note that in the case of $a=0$, Eq.~\eqref{met_kerr} gives the line element in the  Schwarzschild spacetime~\eqref{met_sch}. Let us consider the geodesic motion of a test particle with mass $m$ on the equatorial plane (i.e., $\theta = \pi/2$). The Lagrangian of the particle is given by
\begin{align}
  \mathcal{L} 
  = \frac{1}{2} \1 m \1 \biggl[ - \biggl( 1- \frac{2M}{r} \biggr) \, \dot{t}^2 - \frac{4\1 aM}{r}\,  \dot{t} \, \dot{\phi} + \frac{r^2}{\Delta} \, \dot{r}^2 + \biggl( r^2 + a^2 + \frac{2Ma^2}{r} \biggr) \, \dot{\phi}^2 \biggr] , \label{Kerr_lag}
\end{align}
where the dot denotes the derivative with respect to the proper time of the particle. Associated with the spacetime symmetry, the energy and the angular momentum of the particle
\begin{align}
  E := - \pdv{\mathcal{L}}{\dot{t}} , \quad L := \pdv{\mathcal{L}}{\dot{\phi}} ,
\end{align}
are conserved. Evaluating the derivatives from~\eqref{Kerr_lag} and solving the two equations for $\dot{t}$ and $\dot{\phi}$, we obtain
\begin{align}
  \dot{t} = \frac{( r^3 + a^2 \1 r + 2 M a^2 ) \1 \tilde{E} - 2 \1 a M \tilde{L}}{r \Delta} , \quad \dot{\phi} = \frac{ ( r - 2M ) \, \tilde{L} + 2 \1  a M \tilde{E} }{r \Delta} , \label{kerr_dotphi}
\end{align}
where $\tilde{E} := E/m$ and $\tilde{L} := L/m$. Substituting Eq.~\eqref{kerr_dotphi} into the normalization condition $g_{\mu \nu} \1 \dot{x}^\mu \dot{x}^\nu = -1$, we obtain after some simplification 
\begin{align}
  \dot{r}^2 = \frac{1}{r^3} R(r) , \label{kerr_rdot}
\end{align}
where
\begin{align}
  R(r) := -(1-\tilde{E}^2) \2 r^3 + 2 M r^2 - \Bigl[ \tilde{L}^2 + a^2 (1- \tilde{E}^2) \Bigr] \1  r + 2 M ( \tilde{L}- a \tilde{E})^2.
  \label{9304}
\end{align}
Hereafter, we assume the stable bound orbit and $\tilde{E}^2 < 1$. Then we can find that the function $R(r)$ has three real zeros, which correspond to the turning points of the orbit. Let us express the zeros as $r_0$, $r_p$, and $r_a$ in order from smallest to largest (i.e., $r_0 < r_p < r_a$). Note that the $r_p$ and $r_a$ denote the periapsis and apoapsis, respectively, and the bound orbit exists between these turning points. The function $R(r)$ can be factorized as
\begin{align}
  R(r) = (1-\tilde{E}^2) (r - r_0) (r-r_p)(r_a-r). \label{Rin}
\end{align}
Comparing the coefficients of Eqs.~\eqref{9304} and \eqref{Rin}, we obtain three equations:
\begin{align}
  2 M &= (1-\tilde{E}^2) (r_0 + r_p + r_a) , \label{ELeq2}\\[8pt]
  \tilde{L}^2 + a^2 (1- \tilde{E}^2) &= (1-\tilde{E}^2) \bigl\{r_p \2 r_a + r_0 \1 (r_p+r_a) \bigr\} , \label{ELeq3}\\[8pt]
  2 M ( \tilde{L}- a \tilde{E})^2 &= (1-\tilde{E}^2) \2 r_0 \2 r_p \2 r_a . \label{ELeq4}
\end{align}
Note that we can find $r_0 > 0$ from Eq.~\eqref{ELeq4}. Solving Eqs.~\eqref{ELeq2}, \eqref{ELeq3}, and \eqref{ELeq4} for $\tilde{E}$, $\tilde{L}$, and $r_0$, we obtain
\begin{align}
  \tilde{E} &= \sqrt{ \frac{f(r_p,r_a) \mp 4  \2 a \1 M  \sqrt{2 M r_p r_a (r_p+r_a) \Delta_p \Delta_a}}{\Bigl[ r_p \1 r_a (r_p + r_a)-2M( r_p^2 + r_p \1 r_a + r_a^2)  \Bigr]^2 - 8 \1  a^2 M \1 r_p \1 r_a ( r_p + r_a) } }, \label{tilEne} \\[8pt]
  \tilde{L} &= \frac{-2 \1 a M (r_p \1 r_a + \Delta_p + \Delta_a - a^2) \pm \sqrt{2 M \1 r_p \1 r_a (r_p+r_a)\Delta_p \1 \Delta_a} }{(r_p-2M)(r_a-2M)(r_p+r_a)-2 \1 a^2 M} \,  \tilde{E} , \label{tilL} \\[8pt]
  r_{0} &= \frac{2M}{1- \tilde{E}^2  } - (r_p + r_a), \label{r0proret}
\end{align}
where we have defined as
\begin{align}
  f(r_p,r_a) &:= (r_p+r_a)(r_p-2M)(r_a-2M)\Bigl[ r_p \1 r_a (r_p+r_a)-2M(r_p^2 + r_p \1 r_a + r_a^2) \Bigr] \nonumber \\ 
  &\hspace{30truemm} + 2 \1 a^2 M \Bigl[ 2M ( r_p^2 + 3 \1 r_p \1 r_a + r_a^2)-3 \1 r_p \1 r_a (r_p + r_a) \Bigr] , \\[8pt]
  \Delta_p &:= \Delta(r_p) = r_p^2 -2 M r_p + a^2 , \\[8pt]
  \Delta_a &:= \Delta(r_a) = r_a^2 -2 M r_a + a^2 .
\end{align}
We can see that there are two possible branches in Eqs.~\eqref{tilEne} and~\eqref{tilL}. Hereafter, we refer to the orbit with the upper and lower signs in those equations as the prograde and retrograde orbits, respectively. This is motivated by the fact that the upper and lower sign branches correspond to the prograde and retrograde circular orbits in the case of $e = 0$.

From Eqs.~\eqref{kerr_dotphi}, \eqref{kerr_rdot}, \eqref{Rin}, and \eqref{ELeq4}, we obtain after some algebra
\begin{align}
  \left| \dv{\phi}{r} \right| = \frac{ 1 - 2 M ( 1 - \eta )/r }{ | \1  1 - \eta  \1 | \, (1-2 M /r + a^2/r^2)} \, \frac{1}{r^2 \, \sqrt{2M \bigl(\frac{1}{r_0} - \frac{1}{r} \bigr) } \sqrt{\bigl(\frac{1}{r_p}-\frac{1}{r} \bigr) \bigl(\frac{1}{r}-\frac{1}{r_a} \bigr) } } , \label{gyaku1}
\end{align}
Note that we have defined the non-dimensional parameter $\eta$ as
\begin{align}
  \eta := \frac{a \tilde{E}}{\tilde{L}} = a \1 \frac{(r_p-2M)(r_a-2M)(r_p+r_a)-2 \1 a^2 M}{-2 \1 a M (r_p \1 r_a + \Delta_p + \Delta_a - a^2) \pm \sqrt{2 M \1 r_p \1 r_a (r_p+r_a)\Delta_p \1 \Delta_a} } .
\end{align}
Now, we define the magnitude of the change in $\phi$ as the particle moves from the periapsis to the next periapsis as
\begin{align}
  \delta \phi_\text{K} := 2 \2  \int_{r_p}^{r_a}  \biggl| \dv{\phi}{r} \biggr| \dd{r}  . \label{1228_3}
\end{align}
The periapsis shift per round $\Delta \phi_\text{K}$ is defined by subtracting the contribution of Newtonian gravity $2 \pi$ from $\delta \phi_\text{K}$:
\begin{align}
  \Delta \phi_\text{K} := \delta \phi_\text{K} - 2 \pi .
\end{align}
In addition, let us define the eccentricity $e$, the semi-major axis $d$, and the semi-latus rectum $p$ by using $r_p$ and $r_a$ as follows:
\begin{align}
  e := \frac{r_a - r_p }{ r_a + r_p} , \quad  d := \frac{r_p + r_a}{2} , \quad p := \frac{2 \1 r_p \1 r_a}{r_p + r_a} = d(1-e^2) . \label{kidoupara_teigi1}
\end{align}
Using these orbital parameters, $r_p$ and $r_a$ are written as
\begin{align}
 r_p = d(1-e) = p/(1+e) , \quad r_a = d(1+e) = p/(1-e). \label{kidoupara_teigi2}
\end{align}
Substituting Eq.~\eqref{gyaku1} into Eq.~\eqref{1228_3}, and considering the variable transformation $1/r = (1+ e \sin \chi) / p $, we get
\begin{align}
  \delta \phi_\text{K} = \frac{2}{| \1 1 - \eta \1 | \sqrt{2M \bigl( \frac{1}{r_0} -\frac{1}{p} \bigr)} }\int_{-\pi/2}^{\pi/2} \,   \frac{ 1 - 2 M ( 1 - \eta ) (1+ e\1 \sin \chi)/p }{1 -2 M (1+ e\1 \sin \chi)/p +a^2 (1+ e\1 \sin \chi)^2/p^2 }   \, \frac{\dd{\chi}}{\sqrt{1 - \frac{ e \sin \chi }{ p/r_0 -1}  } } .
  \label{hiseki1}
\end{align}
Here, motivated by Walters' work, we define the new expansion parameter $\gamma$ as
\begin{align}
  \gamma := \frac{e}{p/r_0 - 1} = \frac{\frac{2 \1 e M}{p} \left(1-\frac{a^2}{M p} \right)^2}{1-\frac{6 M}{p} \pm \frac{2 \1 a}{\sqrt{M p}} \sqrt{q} + \frac{a^2}{M p} \left[1 + \frac{\left(7+ e^2\right) M}{p} \right] -\frac{a^4 \left(3-e^2\right)}{M p^3}  } , \label{gamma_def}
\end{align}
where
\begin{align}
 q := \left[ 1 - \frac{2 M (1 + e) }{p} + \frac{a^2 (1 + e)^2}{p^2} \right] \left[ 1 - \frac{2 M (1 - e) }{p} + \frac{a^2 (1 - e)^2}{p^2} \right] .
\end{align}
The parameter $\gamma$ recovers Walters' expansion parameter $\beta$ \eqref{def_beta} in the Schwarzschild limit $a = 0$. Note that the denominator in the middle expression of $\gamma$~\eqref{gamma_def} represents the non-dimensional quantity that vanishes in the limit of the ISCO~\cite{Bardeen:1972fi,Shapiro:1983du}. Rewriting Eq.~\eqref{hiseki1} with $\gamma$, we get
\begin{align}
  \delta \phi_\text{K} = \frac{1 - 2 \1 ( 1-\eta ) M/p }{| \1 1 - \eta \1 | \1  \Delta(p)/p^2 } \frac{ 2 }{ \sqrt{2M (1/r_0 - 1/p) }} \int_{-\pi/2}^{\pi/2} \dd{\chi} \frac{1}{ \sqrt{1 - \gamma \sin \chi } } \frac{ 1 - F \gamma \sin \chi }{1 - G \1 \gamma  \sin \chi + H \gamma^2 \sin^2 \chi } ,  \label{hiseki2}
\end{align}
where
\begin{align}
  F &:= \frac{2 \1 (M/p) ( 1 - \eta ) (p/r_0 -1 )}{ 1 - 2 \1 (1 - \eta) (M/p)} , \quad 
  G := \frac{2 \bigl\{ M/p - (a/p)^2 \bigr\} (p/r_0-1 ) }{1 - 2 M/p + (a/p)^2} , \\[8pt]
  H &:= \frac{(a/p)^2 (p/r_0 - 1)^2}{1 - 2 M/p +(a/p)^2} .
\end{align}
Note that $F = G$ and $H = 0$ in the case of $a=0$, respectively. The $\Delta(p)$ in Eq.~\eqref{hiseki2} is the expression in which $r=p$ is substituted into the function defined below Eq.~\eqref{met_kerr}. The integrand in Eq.~\eqref{hiseki2} can be expanded as
\begin{align}
  \frac{1}{\sqrt{1- \gamma  \1 \sin \chi}} \,  \frac{1-F \1 \gamma  \1 \sin \chi}{ 1 -  G \1 \gamma \1 \sin \chi + H \1 \gamma^2  \sin^2 \chi } = \sum_{m=0}^\infty \sum_{k=0}^\infty A_m \1  B_k \2 \gamma^{m+k}  \sin^{m+k} \chi . \label{tenhou20}
\end{align}
Here, the expansion formula
\begin{align}
  \frac{1}{\sqrt{1-x}} = \sum_{n=0}^\infty A_n \1 x^n \quad (\text{for} \, \, \, |x|<1),  \quad A_n := \frac{(2n)!}{2^{2n} (n!)^2} \quad (n = 0, 1, 2,\cdots) , \label{tenkai_kousiki2}
\end{align}
has been used. We have defined $0! := 1$ and $0^n := \delta^{0}_{\ n}$ for convenience, where $\delta^{i}_{\ j}$ is the Kronecker delta. The coefficients $B_n$ ($n=0,1,2,\cdots$) have been defined as \begin{empheq}[left={B_n :=\empheqlbrace}]{alignat=2}
  & \frac{J_{-}^n \1 (2 F-J_{-}) -J_{+}^n \1 (2 F-J_{+}) }{2^n \1 (J_{+}-J_{-})} \label{eq:kerr_Bn} &\quad &\text{(for \1$a < M$)}, \\[5pt]
  & \frac{G^{n-1} \bigl[G + n (G - 2 F) \bigr] }{2^n } \label{eq:kerr_Bn_ex} &       &\text{(for \1$a = M$)},
\end{empheq}
where $J_{\pm} := G \pm \sqrt{G^2-4H}$. We need to use $B_n$ separately in the non-extremal case (i.e., $a < M$) and the extremal case (i.e., $a =M$). We can confirm that Eq.~\eqref{eq:kerr_Bn_ex} is properly recovered as the continuous limit $a \to M$ in Eq.~\eqref{eq:kerr_Bn}. Note that in the case of $a=0$, Eq.~\eqref{eq:kerr_Bn} becomes $B_n = \delta^{0}_{\ n}$ since $F=G$ and $H=0$.

Focusing on Eq.~\eqref{tenhou20}, the infinite sums run independently for $k$ and $m$, and this makes it difficult to find the power of $\gamma$. Therefore, let us rewrite these two independent infinite sums to make the power of $\gamma$ easier to find. Rewriting the right-hand side of Eq.~\eqref{tenhou20} by using the formula
\begin{align}
  \sum_{m=0}^\infty \sum_{k=0}^\infty a_{m,k} \2 x^{m+k}=  \sum_{n=0}^\infty \sum_{l=0}^n a_{l,n-l} \2 x^{n} , \label{sum_kousiki1}
\end{align}
and putting it back in Eq.~\eqref{hiseki2}, we obtain
\begin{align}
  \delta \phi_\text{K} = \frac{1 - 2 \1 ( 1-\eta ) M/p }{| \1 1 - \eta \1 | \1 \Delta(p)/p^2 } \frac{ 2 }{ \sqrt{2M (1/r_0 - 1/p) }} \sum_{n=0}^\infty \sum_{k=0}^{2n} A_{k} \1 B_{2n-k} \1 \gamma^{2n} \int_{-\pi/2}^{\pi/2} \dd{\chi} \1 \sin^{2n} \chi .  \label{tenhou18}
\end{align}
Note that the odd powers of $\sin \chi$ do not contribute to the integral. Then, using the integral formula
\begin{align}
  \int_{-\pi/2}^{\pi/2} \1 \sin^{2n} d\chi  = A_{n}  \1  \pi \quad  (n=0,1,2,\cdots), \label{sekibun_kousiki3}
\end{align}
where $A_n$ is given by Eq.~\eqref{tenkai_kousiki2}, Eq.~\eqref{tenhou18} can be integrated as
\begin{align}
 \delta \phi_\text{K} = \frac{1 - 2 \1 ( 1-\eta ) M/p }{ | \1 1 - \eta \1 | \1 \Delta(p)/p^2 } \frac{ 2 \pi }{ \sqrt{2M (1/r_0 - 1/p) }} \sum_{n=0}^\infty \Biggl( \1 \sum_{k=0}^{2n} A_{n} \1 A_{k}  \1 B_{2n-k}  \Biggr) \1 \gamma^{2n} .  \label{tenhou21}
\end{align}
Finally, we obtain a new series representation for the periapsis shift per round in the Kerr spacetime:
\begin{align}
  \Delta \phi_\text{K} = 2 \pi \left[ \frac{1 - 2 \1 ( 1 - \eta ) M/p }{| \1 1 - \eta \1 | \1 \Delta(p)/p^2 } \frac{ 1 }{ \sqrt{2M (1/r_0 - 1/p) }} \sum_{n=0}^\infty \Biggl( \1 \sum_{k=0}^{2n} A_{n} \1 A_{k}  \1 B_{2n-k}  \Biggr) \1 \gamma^{2n} -1 \right].
  \label{tenhou21-4}
 \end{align}

In the case of $a=0$ in Eq.~\eqref{tenhou21-4}, the summation for $k$ only leaves $k=2n$ terms since $B_{2n-k}=\delta^{0}_{\ 2n-k}$, and we confirm that Walters' representation which is for the Schwarzschild spacetime~\eqref{sch_kin_revolution} is recovered. In addition, expanding Eq.~\eqref{tenhou21-4} in powers of $M/p$ and $a/p$, we get the PN expansion formula (e.g., \cite{boyer_price_1965,Bini:2005dy,Vogt:2008zs,de2011estimating,He:2023joa}). The specific expression of it with higher order terms is presented in Appendix~\ref{ap:kin_higher}. Furthermore, in the case of $e = 0$, Eq.~\eqref{tenhou21-4} provides an exact formula for the periapsis shift of the quasi-circular orbit
\begin{align}
  \Delta \phi_{\text{K,qc}} &= 2\1 \pi \left[ \frac{1}{ \sqrt{1 - 6 M/d \pm 8 \1 a \1 M^{\frac{1}{2}} / d^{\frac{3}{2}} - 3 \1 a^2/d^2 } }  -1 \right] .
  \label{Kerrzyunen1} 
\end{align}
As in the case of the Schwarzschild spacetime~\eqref{sch_peri_shift_qc}, this formula diverges in the limit of the ISCO. This divergence reflects the fact that there is an apoapsis but no periapsis in the vicinity of ISCO, which is usually at the inflection point of the effective potential.

\subsection{Behavior of the expansion parameter}\label{subsec:behave_kerr_para}
\begin{figure}[htbp]
  \begin{minipage}{0.45\linewidth}
    \centering
    \includegraphics[keepaspectratio, scale=0.108]{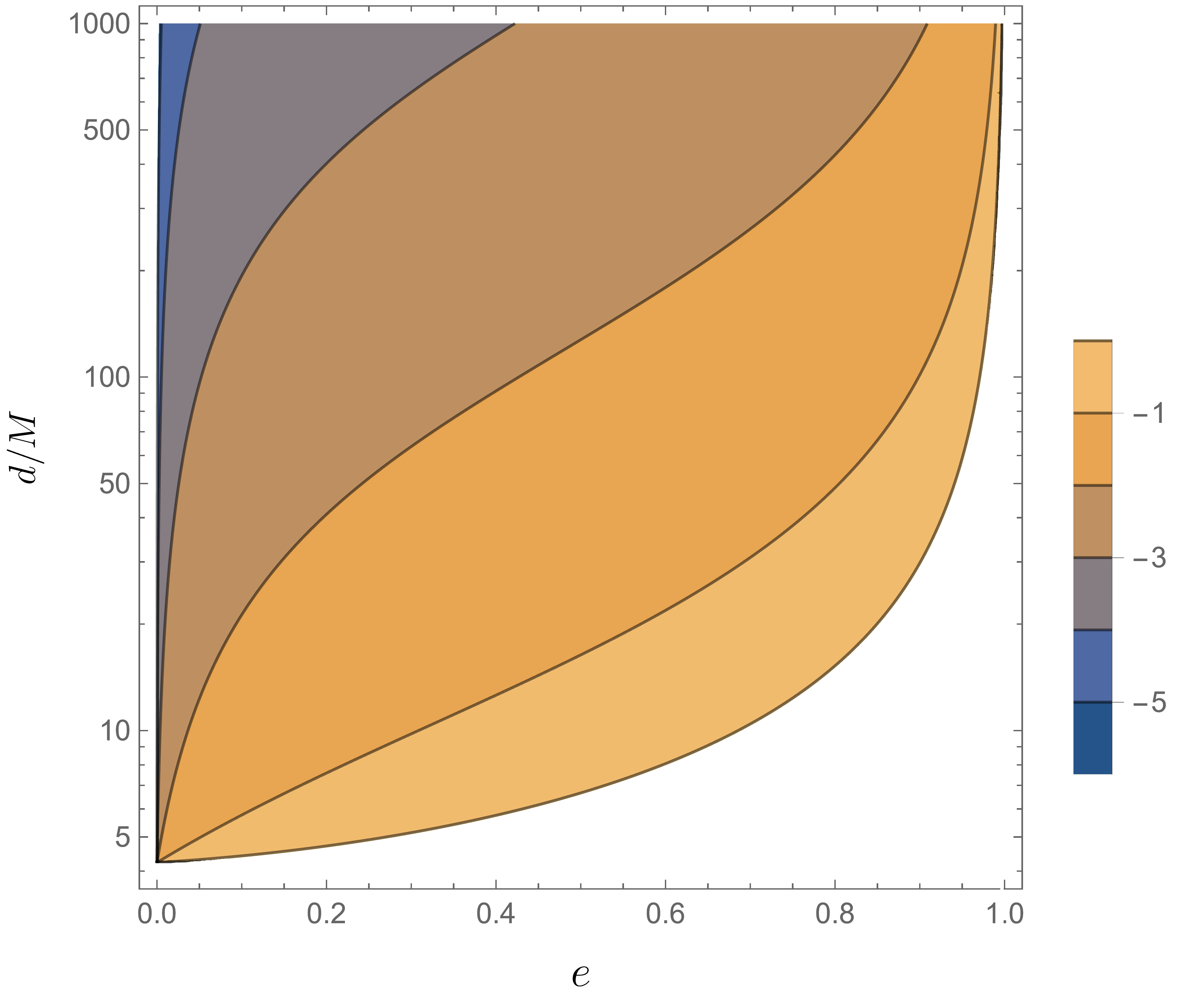}
    \subcaption{New expansion parameter}
    \label{fig:para_kerr_para_contour_a}
  \end{minipage} 
  \hspace{8truemm}
  \begin{minipage}{0.45\linewidth}
    \centering
    \includegraphics[keepaspectratio, scale=0.45]{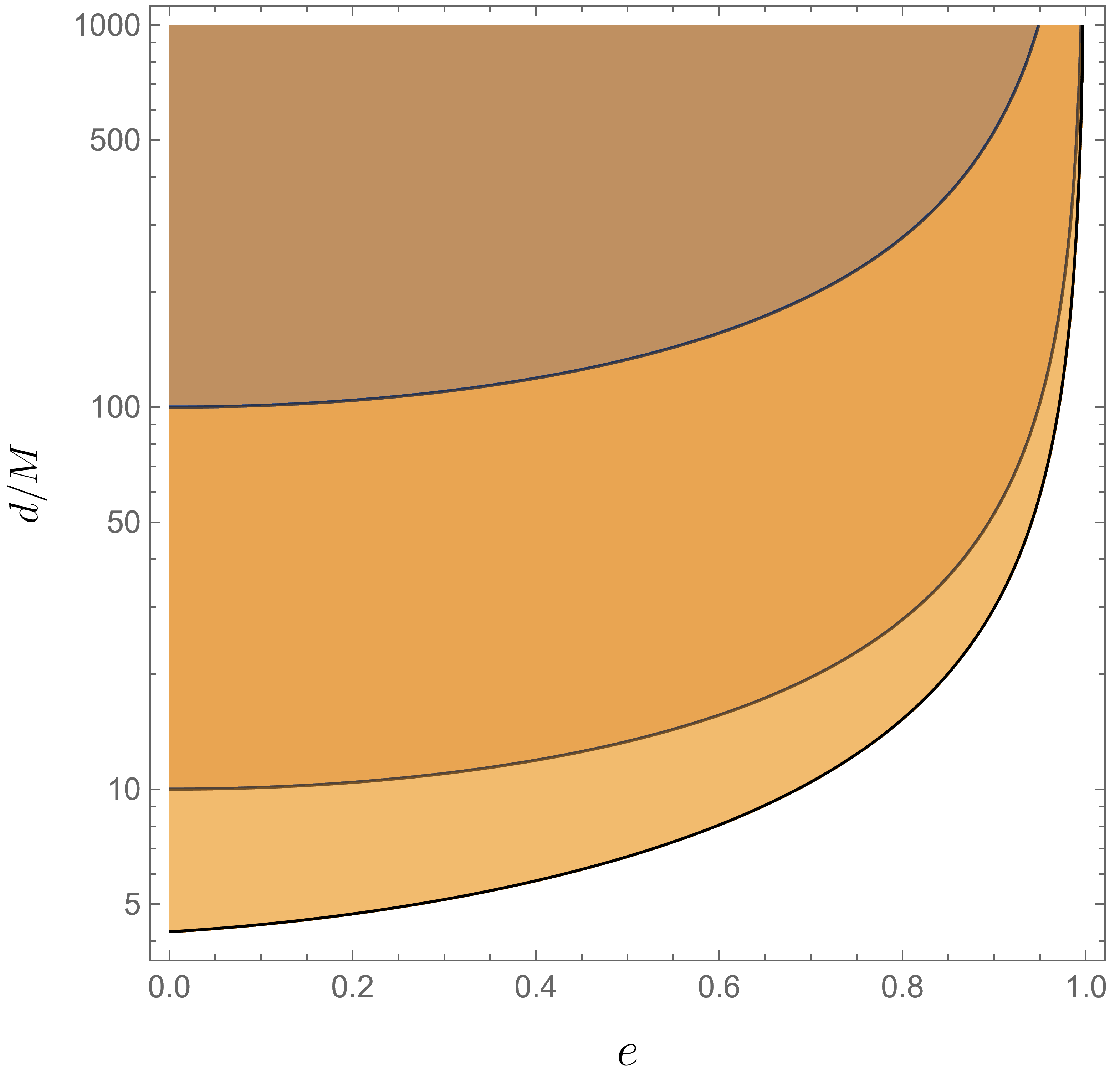}
    \subcaption{PN expansion parameter}
    \label{fig:para_kerr_para_contour_b}
  \end{minipage}
  \caption{Contour plots of the new expansion parameter $\gamma$ and the PN expansion parameter $M/p=M/[d(1-e^2)]$. Base 10 logarithms of the parameters are plotted. In the case of $e \to 0$, $\gamma$ approaches zero, while the PN expansion parameter approaches a non-zero value. It can be seen that the larger $d/M$ also leads to the smaller both parameters. We only plot the prograde case here. The Kerr parameter is fixed to be $a/M=1/2$. Note that the parameter regions that do not satisfy the conditions $r_p > r_0$, $r_p > r_+$, and $r_a > r_\text{ISCO}$ are filled in white, where $r_+:=M+\sqrt{M^2-a^2}$ and $r_\text{ISCO}$ denote the radii of the horizon and the ISCO~\cite{Bardeen:1972fi,Shapiro:1983du}, respectively.}
  \label{fig:para_kerr_para_contour}
\end{figure}

Let us discuss the dependence of the new expansion parameter $\gamma$ on the orbital parameters. We show the contour plots of $\gamma$ and the PN expansion parameter $M/p=M/[d(1-e^2)]$ in Fig.~\ref{fig:para_kerr_para_contour}. We can see that both parameters decrease as $e$ is decreased. It is noteworthy that the major difference in dependence on $e$ appears at $e \to 0$. In this case, $\gamma$ approaches zero since it is proportional to $e$ (see Eq.~\eqref{gamma_def}), while the PN expansion parameter approaches a non-zero value $M/d$. It can be seen that the larger $d/M$ also leads to the smaller $\gamma$ and the PN expansion parameter.

\begin{figure}[htbp]
\hspace{-8truemm}
  \begin{minipage}{0.45\linewidth}
    \centering
    \includegraphics[keepaspectratio, scale=0.45]{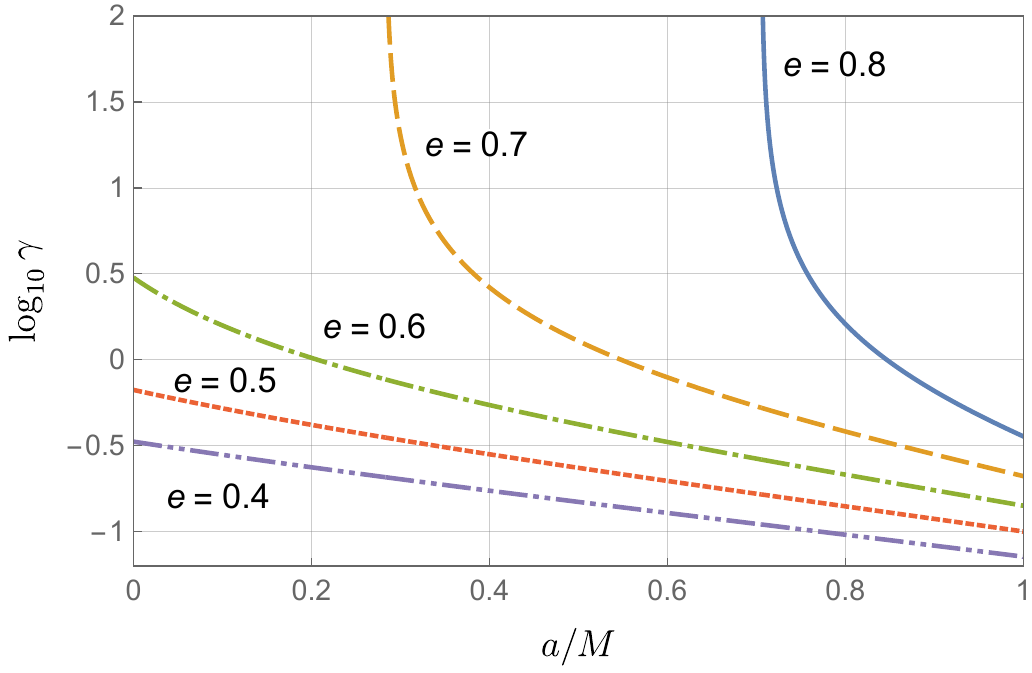}
    \subcaption{prograde}
    \label{fig:Kerr_para_gamma_aplot_a}
  \end{minipage} 
  \hspace{7truemm}
  \begin{minipage}{0.45\linewidth}
    \centering
    \includegraphics[keepaspectratio, scale=0.45]{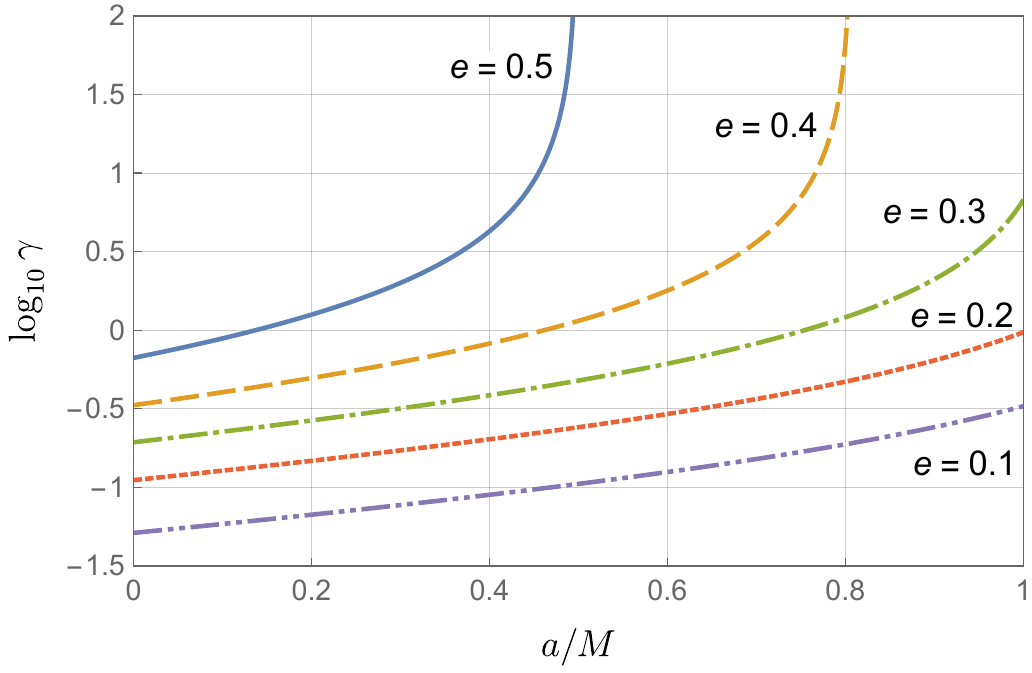}
    \subcaption{retrograde}
    \label{fig:Kerr_para_gamma_aplot_b}
  \end{minipage}
  \caption{Dependence of the new expansion parameter $\gamma$ on the Kerr parameter $a$. The prograde (a) and retrograde (b) cases are shown. The semi-major axis is fixed to be $d/M=10$. Base 10 logarithm of $\gamma$ is plotted. We can see that the $\gamma$ monotonically decreases with increasing $a/M$ in the prograde case. In contrast, it can be seen that the $\gamma$ monotonically decreases with decreasing $a/M$ in the retrograde case.}
  \label{fig:Kerr_para_gamma_aplot}
\end{figure}

The dependence of $\gamma$ on the Kerr parameter $a$ is also shown in Fig.~\ref{fig:Kerr_para_gamma_aplot}. From Fig.~\ref{fig:Kerr_para_gamma_aplot_a}, we can see that the $\gamma$ monotonically decreases with increasing $a/M$ in the prograde case. In contrast, from Fig.~\ref{fig:Kerr_para_gamma_aplot_b}, it can be seen that the $\gamma$ monotonically decreases with decreasing $a/M$ in the retrograde case. This property can be explained by the analytical form of $\gamma$~\eqref{gamma_def}. One can see that the strongest contribution of $a/M$ in the expression of $\gamma$ is the first order term of $a/M$ in the denominator. The sign of the term depends on whether the orbit is prograde or retrograde. If the orbit is prograde, the sign is positive, and the larger $a/M$ leads to the smaller $\gamma$. On the other hand, if the orbit is retrograde, the sign is negative, and the smaller $a/M$ leads to the smaller $\gamma$. We can also understand this more intuitively as follows. The $\gamma$ denotes how eccentric the orbit is and how close it is to the ISCO. For $a=0$ the radius of ISCO is $r/M=6$ in both the prograde and retrograde case, but as $a/M$ is increased from 0 to 1, the ISCOs for prograde and retrograde go far from and close to the orbit at $d/M=10$, respectively.

\subsection{Accuracy of the truncated new representation}
Here, let us investigate the accuracy of the formula that is obtained by truncating the new series representation up to a finite number of terms, and compare it with that of the truncated PN expansion formula. Now, we define the periapsis shift per round as the integral form:
\begin{align}
  I = \frac{1 - 2 \1 ( 1-\eta ) M/p }{| \1 1 - \eta \1 | \1 \Delta(p)/p^2 } \frac{ 2 }{ \sqrt{2M (1/r_0 - 1/p) }} \int_{-\pi/2}^{\pi/2} \frac{\dd{\chi}}{\sqrt{1 - \gamma \sin \chi } } \, \frac{ 1 - F \gamma \sin \chi }{1 - G \1 \gamma  \sin \chi + H \gamma ^2 \sin ^2 \chi }  - 2 \pi . \label{I_integrate} 
\end{align}
We can calculate the amount of the periapsis shift by integrating Eq.~\eqref{I_integrate} numerically. Let us also truncate the infinite series of new series representation~\eqref{tenhou21-4} up to $n=N-1$ ($N = 1,2,3, \cdots$) as
\begin{align}
  \Delta \phi_\text{K} (N) := 2 \pi \left[ \frac{1 - 2 \1 ( 1-\eta ) M/p }{| \1 1 - \eta \1 | \1 \Delta(p)/p^2 } \frac{ 1 }{ \sqrt{2M (1/r_0 - 1/p) }} \sum_{n=0}^{N-1} \Biggl(\,  \sum_{k=0}^{2n} A_{n} \1 A_{k}  \2 B_{2n-k}  \Biggr) \1 \gamma^{2n}  - 1 \right] .
  \label{newform_yuugen}
\end{align}
For example, the formulae with the first term and the first two terms taken in the series are written as
\begin{align}
  \Delta \phi_\text{K} (1) &= 2 \pi \left[ \frac{1 - 2 \1 ( 1-\eta ) M/p }{| \1 1 - \eta \1 | \1 \Delta(p)/p^2 } \frac{ 1 }{ \sqrt{2M (1/r_0 - 1/p) }} - 1 \right] , \\[8pt]
  \Delta \phi_\text{K} (2) &= 2 \pi \Biggl[ \frac{1 - 2 \1 ( 1-\eta ) M/p }{| \1 1 - \eta \1 | \1 \Delta(p)/p^2 } \frac{ 1 }{ \sqrt{2M (1/r_0 - 1/p) }} \nonumber \\
  & \hspace{30truemm} \times \left\{ 1  + \frac{ 3 +4 \1 G +8 H  -4 F (1 + 2 \1 G) +8 \1 G^2  }{16}\, \gamma ^2 \right\} - 1 \Biggr] , \label{newform_1}
\end{align}
respectively. We also define the PN expansion formulae taken up to the $N$-th order:
\begin{align}
  \Delta \phi_\text{K,PN} (N) := \sum_{n=0}^{2(N-1)} \mathcal{B}_{1+n/2} \quad (N=1,\2 3/2,\2 2,\2 5/2,\cdots) , \label{kerr_PN_kyuusuu}
\end{align}
where $\mathcal{B}_{k} \, \,  (k=1,\2 3/2,\2 2, \2 5/2, \cdots)$ is the $k$\1-\1th order term of $M/p$ and $a/p$ in the PN expansion formula, for example,
\begin{align}
  \mathcal{B}_{1} := \frac{6 \1 \pi M}{p} , \quad \mathcal{B}_{3/2} := \mp \2 \frac{8 \3 \pi \3 a \3 M^{1/2}}{p^{3/2}} , \quad  \mathcal{B}_{2} := \frac{3 \1 \pi \1 (18+e^2) M^2}{2 \1 p^2} +  \frac{3 \1 \pi \1 a^2}{p^2} . \label{Kerr_PN_Bn}
\end{align}
See Eq.~\eqref{Kerr_PN_higher} in Appendix~\ref{ap:kin_higher} for higher order terms. Here, let us define the relative errors of the truncated new representation and the truncated PN expansion formula to the numerical integration of Eq.~\eqref{I_integrate} as
\begin{align}
  \Delta_\text{K,NR}(N) := \biggl| \2 \frac{\Delta \phi_\text{K} (N)}{I} -1 \2 \biggr| , \quad 
  \Delta_\text{K,PN} (N) := \biggl| \2 \frac{\Delta \phi_\text{K,PN} (N) }{I} -1 \2 \biggr| ,
\end{align}
respectively.

\begin{figure}[htbp]
\hspace{-10truemm}
  \begin{minipage}{0.45\linewidth}
    \centering
    \includegraphics[keepaspectratio, scale=1.09]{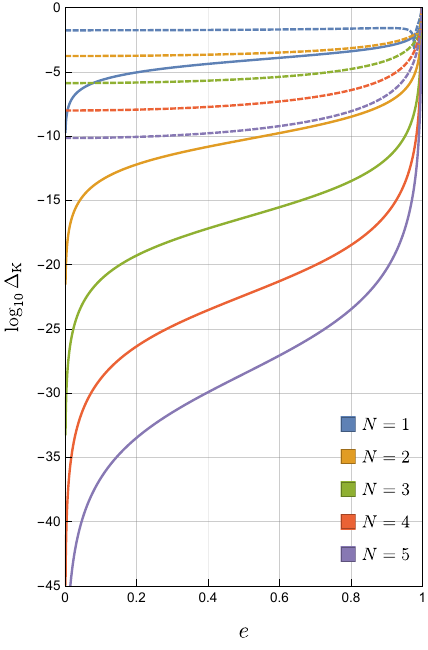}
    \subcaption{$d/M=1000$}
    \label{fig:Kerrgosaeplotd1000}
  \end{minipage} 
  \hspace{7truemm}
  \begin{minipage}{0.45\linewidth}
    \centering
    \includegraphics[keepaspectratio, scale=1.09]{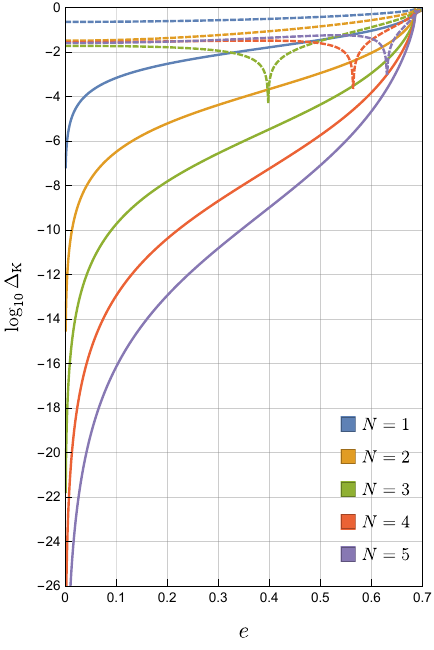}
    \subcaption{$d/M=10$}
    \label{fig:Kerrgosaeplotd10}
  \end{minipage}
  \caption{Dependence of the relative errors of the truncated new representation $\Delta_\text{K,NR}(N)$ and the truncated PN expansion formula $\Delta_\text{K,PN}(N)$ on the eccentricity $e$. Solid and dashed lines represent the relative errors $\Delta_\text{K,NR}(N)$ and $\Delta_\text{K,PN}(N)$, respectively. In the case of $e \to 0$, it can be seen that the errors of the truncated new representation approach zero, while the errors of the truncated PN expansion formula approach non-zero values. We only plot the prograde case and fix the non-dimensional Kerr parameter as $a/M=1/2$ here.}
  \label{fig:Kerrgosaeplot}
\end{figure}

\begin{figure}[htbp]
\hspace{-10truemm}
  \begin{minipage}{0.45\linewidth}
    \centering
    \includegraphics[keepaspectratio, scale=1.09]{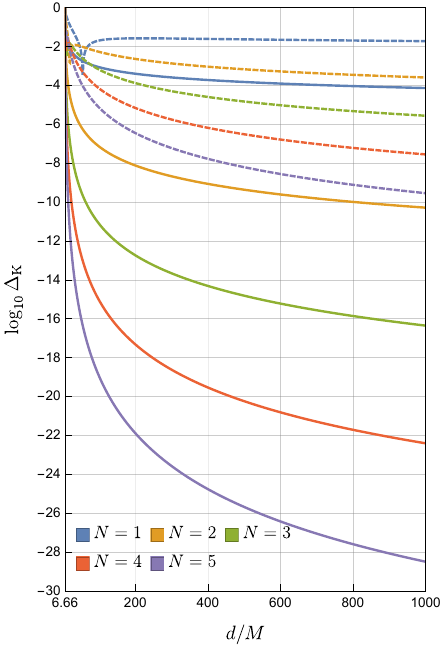}
    \subcaption{$e=1/2$}
    \label{fig:Kerrgosadplote05}
  \end{minipage} 
  \hspace{7truemm}
  \begin{minipage}{0.45\linewidth}
    \centering
    \includegraphics[keepaspectratio, scale=1.09]{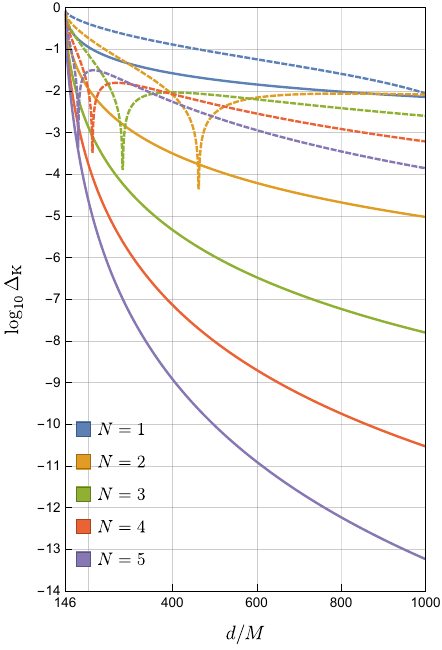}
    \subcaption{$e=0.98$}
    \label{fig:Kerrgosadplote098}
  \end{minipage}
  \caption{Dependence of the relative errors of the truncated new representation $\Delta_\text{K,NR}(N)$ and the truncated PN expansion formula $\Delta_\text{K,PN}(N)$ on the semi-major axis $d$. Solid and dashed lines represent the relative errors $\Delta_\text{K,NR}(N)$ and $\Delta_\text{K,PN}(N)$, respectively. It can be seen that the errors of the truncated new representation decrease as $d/M$ is increased. We only plot the prograde case and fix the Kerr parameter as $a/M=1/2$ here. The origin of the horizontal axis is determined by using the condition $r_p > r_0$.}
  \label{fig:Kerrgosadplot}
\end{figure}

\begin{figure}[htbp]
\centering
\begin{minipage}{0.32\linewidth}
    \centering
    \includegraphics[keepaspectratio, scale=1.2]{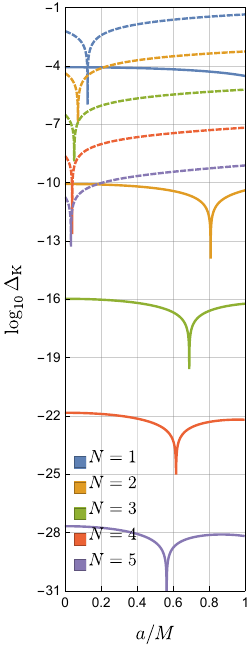}
    \subcaption{$d/M=1000$, prograde}
    \label{fig:Kerrgosaaplotd1000pro}
\end{minipage}
\begin{minipage}{0.32\linewidth}
    \centering
    \includegraphics[keepaspectratio, scale=1.2]{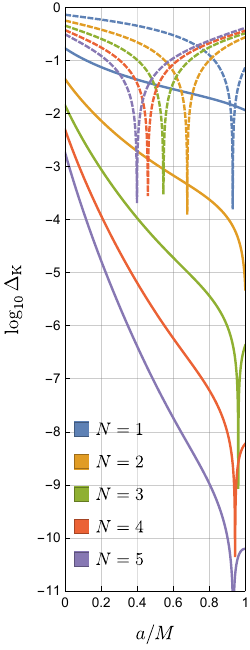}
    \subcaption{$d/M=10$, prograde}
    \label{fig:Kerrgosaaplotd10pro}
\end{minipage}
\begin{minipage}{0.32\linewidth}
    \centering
    \includegraphics[keepaspectratio, scale=1.2]{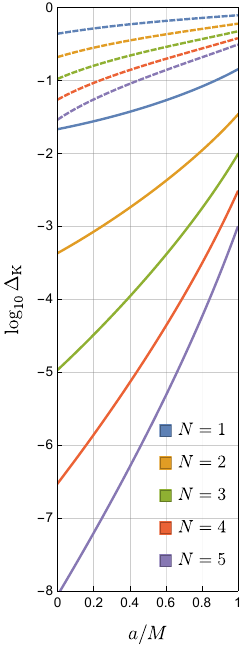}
    \subcaption{$d/M=15$, retrograde}
    \label{fig:Kerrgosaaplotd15ret}
\end{minipage}
\caption{Dependence of the relative errors of the truncated new representation $\Delta_\text{K,NR}(N)$ and the truncated PN expansion formula $\Delta_\text{K,PN}(N)$ on the Kerr parameter $a$. Solid and dashed lines represent the errors $\Delta_\text{K,NR}(N)$ and $\Delta_\text{K,PN}(N)$, respectively. The eccentricity is fixed as $e=1/2$.}
\label{fig:Kerrgosaaplot}
\end{figure}

Figs.~\ref{fig:Kerrgosaeplot}, ~\ref{fig:Kerrgosadplot}, and \ref{fig:Kerrgosaaplot} show the dependence of the relative errors of the truncated new representation $\Delta_\text{K,NR}(N)$ and the truncated PN expansion formula $\Delta_\text{K,PN}(N)$ on the eccentricity $e$, the semi-major axis $d$, and the Kerr parameter $a$, respectively. First, we focus on the dependence on the eccentricity $e$. From Fig.~\ref{fig:Kerrgosaeplotd1000}, we can see that the relative error of the truncated new representation (solid lines)  decreases as $e$ is decreased. It can also be seen that the error decreases as $N$ is increased. It is noteworthy that the error of the truncated new representation rapidly approaches zero in the case of $e \to 0$. Dashed lines in Fig.~\ref{fig:Kerrgosaeplotd1000} show the relative error of the truncated PN expansion formula for comparison. We can see that the error of the truncated PN expansion also decreases as $e$ is decreased and $N$ is increased. We emphasize that the errors of the truncated PN expansion formula approach to non-zero values in the case of $e \to 0$ in Fig.~\ref{fig:Kerrgosaeplotd1000}. This is the major difference in the dependence of the accuracy on $e$ between the truncated new representation and the truncated PN expansion formula. This difference would be due to the difference in the dependence of the expansion parameters on $e$ as discussed in Sec.~\ref{subsec:behave_kerr_para}. Furthermore, Fig.~\ref{fig:Kerrgosaeplotd10} shows the dependence of the accuracy on $e$ when the orbit is closer to the center ($d/M=10$). It can be seen that the truncated new representation seems to converge with smaller $e$ (solid lines), while the truncated PN expansion formula is poorly converged (dashed lines). Note that the truncated new representation seems to break down at $e \sim 0.7$. It is the advantage that if $e$ is sufficiently small, the truncated new representation has higher accuracy than the truncated PN expansion formula even if the orbit is closer to the center, where the convergence of the PN expansion formula is not guaranteed.

Next, we discuss the dependence of the relative errors on the semi-major axis $d$. From Fig.~\ref{fig:Kerrgosadplote05}, we can see that the error of the truncated new representation (solid lines) decreases as $d/M$ is increased. Dashed lines in Fig.~\ref{fig:Kerrgosadplote05} show the dependence of the errors of the truncated PN expansion formula on $d/M$ for comparison. We can see that the larger $d/M$ results in the higher accuracy of the truncated PN expansion formula as the truncated new representation. We emphasize that the truncated new representation has higher accuracy than the truncated PN expansion formula if we compare for the same $N$ in Fig.~\ref{fig:Kerrgosadplote05}. In addition, Fig.~\ref{fig:Kerrgosadplote098} shows the dependence of the errors on $d/M$ in the case where the orbit is highly eccentric ($e=0.98$). We can see that the error of the truncated new representation decreases as $d/M$ is increased. The larger $N$ also leads to the higher accuracy. One of the advantages of the truncated new representation is that even if the orbit is highly eccentric, it has higher accuracy than the truncated PN expansion formula if $d/M$ and $N$ are sufficiently large.

Finally, let us discuss the dependence of the relative errors on the Kerr parameter $a$. Fig~\ref{fig:Kerrgosaaplotd1000pro} shows the dependence of the errors of the truncated new representation and the truncated PN expansion formula on the Kerr parameter $a$ in the case of $d/M=1000$. We can see that the error of the truncated PN expansion formula decreases as $a/M$ is decreased in the case of at least $a/M > 0.2$. On the other hand, the dependence of the error of the truncated new representation is less visible. In other words, in case the orbit is far from the center, the effect of the BH spin does not contribute to the new representation as strongly as the eccentricity $e$. Now, let us consider the case where the orbit is closer to the center. Figs.~\ref{fig:Kerrgosaaplotd10pro} and~\ref{fig:Kerrgosaaplotd15ret} show the dependence of the relative errors on $a/M$ in the prograde case with $d/M=10$ and in the retrograde case with $d/M=15$, respectively. Fig.~\ref{fig:Kerrgosaaplotd10pro}, we can see that the larger $a/M$ results in the higher accuracy of the truncated new representation in the prograde case. In contrast, from Fig.~\ref{fig:Kerrgosaaplotd15ret}, the smaller $a/M$ leads to the higher accuracy of the truncated new representation in the retrograde case. This property would be caused by the dependence of the new expansion parameter $\gamma$ on $a/M$ as discussed in Sec.~\ref{subsec:behave_kerr_para}. Dashed lines in Figs.~\ref{fig:Kerrgosaaplotd10pro} and~\ref{fig:Kerrgosaaplotd15ret} show the errors of the truncated PN expansion formula. It can be seen that it is uncertain whether the PN expansion formula converges or not in the prograde case with $d/M=10$. On the other hand, the error of the truncated PN expansion formula decreases as $a/M$ is decreased in the retrograde case with $d/M=15$. If we consider even higher order terms, the PN expansion formula could converge in this retrograde case. One of the advantages of the truncated new representation is that even if the orbit is close to the center where the convergence of the PN expansion formula is not guaranteed, it has higher accuracy than the truncated PN expansion formula as long as $N$ is large, $a/M$ is large in the prograde case, and $a/M$ is small in the retrograde case.

\section{New series expansion method for the periapsis shift in general spacetimes}
\label{sec:new_method}
\subsection{General method}
In this section, we propose a new general series expansion method for the periapsis shift in various spacetimes by generalizing Walters' work. The method has four steps as follows:
\begin{enumerate}
  \item[(I)] Obtain the integral that represents the periapsis shift. 
  \item[(II)]
  Expand the integrand in powers of the eccentricity and find the zeroth order term. Factor out it from the integrand and expand the remaining part in powers of the eccentricity again.
  \item[(III)]
  Focusing on each term, define the expansion parameter as the eccentricity divided by a non-dimensional quantity that vanishes in the limit of the ISCO, and expand the integrand in powers of the parameter.
  \item[(IV)]
  By integrating it, obtain a new series representation for the periapsis shift.
\end{enumerate}
As we will see later, the new series expansion of the periapsis shift in the Kerr spacetime which we have proposed in the previous section is actually a specific application of the general method. As another specific example, we consider the application of the general method to the Chazy-Curzon spacetime and obtain a new series representation for the periapsis shift. 
\subsection{Application to the Chazy-Curzon spacetime}
\subsubsection{Chazy-Curzon solution}
Before we consider the application of the new method to the Chazy-Curzon spacetime, we briefly review this spacetime. Axisymmetric, static, and vacuum solutions of Einstein's field equations can be described by the Weyl metric~\cite{Weyl:1917,Weyl:2012} (see also Ref.~\cite{Griffiths:2009dfa}). Using cylindrical coordinates $(t,\rho,z,\phi)$ with $-\infty < t < \infty$, $\rho \geq 0$, $-\infty < z < \infty$, and $0 \leq \phi < 2 \pi$, the line element is written as
\begin{align}
  \dd{s}^2 = - \mathrm{e}^{2 \Phi} \dd{t}^2 + \mathrm{e}^{2(\Lambda - \Phi)} ( \dd{\rho}^2 + \dd{z}^2 ) + \rho^2 \mathrm{e}^{-2 \Phi} \dd{\phi}^2, \label{CC_met}
\end{align}
where the metric functions $\Phi$ and $\Lambda$ depend on $\rho$ and $z$. The vacuum Einstein's field equations imply
\begin{align}
    \Phi_{,\rho \rho} + \frac{1}{\rho} \1 \Phi_{,\rho} + \Phi_{,z z} = 0 , \quad
    \Lambda_{,\rho} - \rho \left( \Phi_{,\rho}^2 - \Phi_{,z}^2 \right) = 0 , \quad
    \Lambda_{,z} - 2 \1 \rho \2 \Phi_{,\rho} \1 \Phi_{,z} = 0. \label{CC_eqs}
\end{align}
The first equation can be recognized as the simple three-dimensional Laplace equation in the cylindrical coordinates. Therefore $\Phi$ can be regarded as the analogy of the Newtonian potential. If we find a solution $\Phi$ of the first equation, then we can also find a solution $\Lambda$ by integrating the second and third equations. The simplest asymptotically flat Weyl solution is generated by the Newtonian potential of a spherically symmetric point mass with mass $M$, which is located at the origin of the cylindrical coordinate $(\rho,z)=(0,0)$:
\begin{align}
    \Phi = - \frac{M}{R}, \quad  \Lambda = - \frac{M^2 \rho^2}{2 R^4}, \quad  R:= \sqrt{\rho^2 + z^2}, \label{cc_sol}
\end{align}
where the constant of the integration for $\Lambda$ has been set to zero to satisfy the regularity condition along the $z$-axis
\begin{align}
    \lim_{\rho \to 0} \Lambda = 0.
\end{align}
The solution~\eqref{cc_sol} which is generated by the monopole potential is the Chazy-Curzon solution~\cite{Chazy:1924,Curzon:1925}. The source of this spacetime is sometimes referred to as the Chazy-Curzon particle. There is a curvature singularity at the origin which is not surrounded by an event horizon and is therefore a naked singularity.
\subsubsection{Innermost stable circular orbit}
\begin{figure}[hbtp]
  \centering
  \includegraphics[scale=1.1]{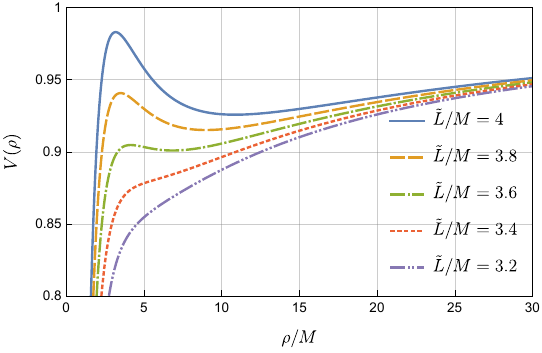}
  \caption{Plot of the effective potential $V(\rho)$ for the timelike geodesic motion in the Chazy-Curzon spacetime. The larger angular momentum gives rise to a higher potential barrier. We can get a general picture of the orbits by considering the magnitude relationship between $V(\rho)$ and $\tilde{E}^2$, as we often do when discussing the orbits in the Schwarzschild spacetime.}
  \label{fig:CC_ep}
\end{figure}

Let us consider the geodesic motion of a test particle with mass $m$ which is bound in the $z=0$ plane. The Lagrangian of the particle is written as
\begin{align}
  \mathcal{L} = \frac{1}{2} \1 m \left[ - \mathrm{e}^{2 \Phi} \dot{t}^2 + \mathrm{e}^{2(\Lambda - \Phi)}  \dot{\rho}^2  + \rho^2 \mathrm{e}^{-2 \Phi} \dot{\phi}^2  \right].
\end{align}
Since the spacetime symmetry, the energy and the angular momentum of the particle
\begin{align}
  E := - \pdv{\mathcal{L}}{\dot{t}} = m \1 \mathrm{e}^{2 \Phi} \dot{t} , \quad 
  L := \pdv{\mathcal{L}}{\dot{\phi}} = m \1 \rho^2 \mathrm{e}^{-2\Phi} \dot{\phi} ,
\end{align}
are conserved. From these equations, one can find
\begin{align}
 \dot{t} = \tilde{E} \1 \mathrm{e}^{-2 \Phi} , \quad \dot{\phi} = \frac{\tilde{L} \1 \mathrm{e}^{2 \Phi}}{\rho^2}, \label{CC_phidot}
\end{align}
where $\tilde{E} := E/m$ and $\tilde{L} := L/m$. Then, the normalization condition gives the equation of the radial motion
\begin{align}
  \dot{\rho}^2 = \mathrm{e}^{M^2/\rho^2} \left[  \tilde{E}^2 - V(\rho) \right], \label{CC_rhodot} 
\end{align}
where we have defined the effective potential as
\begin{align}
  V(\rho) := \mathrm{e}^{- 2 M / \rho } + \frac{\tilde{L}^2 \1 \mathrm{e}^{- 4 M / \rho}}{\rho^2} . \label{CC_effective_pottential_V}
\end{align}
By virtue of $\mathrm{e}^{M^2/\rho^2}$ always being positive, we can get a general picture of the orbits by considering the magnitude relationship between $V(\rho)$ and $\tilde{E}^2$, as we often do when discussing the orbits in the Schwarzschild spacetime. Fig.~\ref{fig:CC_ep} shows the plot of $V(\rho)$. From Fig.~\ref{fig:CC_ep}, it can be seen that the larger angular momentum gives rise to a higher potential barrier. 

Now, let us focus on circular orbits. The conditions for circular orbits $\dot{\rho} = 0$ and $\ddot{\rho} = 0$ yield $\tilde{E}^2 = V(\rho)$ and $V^\prime(\rho) = 0$, respectively (see Appendix~\ref{ap:CC_circular_condition}), where the prime denotes the derivative with respect to $\rho$. From these equations, we can find the energy and the angular momentum in circular orbits are given by
\begin{align}
\tilde{E}^2 = \frac{\mathrm{e}^{-2M/\rho} (\rho - M )}{\rho - 2 M } , \quad \tilde{L}^2 = \frac{M \rho^2 \mathrm{e}^{2M/\rho}}{\rho - 2 M },
\end{align}
respectively. In addition, the condition for stability of circular orbits $V^{\prime \prime} (\rho) \geq 0$ (see Appendix~\ref{ap:CC_circular_condition}) implies 
\begin{align}
\rho^2 -6 M \rho + 4 M^2 \geq 0 . \label{cc_eq_isco}
\end{align}
Solving a quadratic equation in $\rho$ for the limiting case of equality, we can get a solution
\begin{align}
  \rho_\text{ISCO} = (3 + \sqrt{5} )M ,
\end{align}
which denotes the radius of the ISCO~\cite{Gonzalez:2011fb}.

\subsubsection{New series representation for the periapsis shift}
Let us consider the application of the general method to the Chazy-Curzon spacetime and obtain a new series representation for the periapsis shift. First, we obtain the integral that represents the periapsis shift (step~(I)). In order to keep a connection with the work~\cite{Bini:2005dy}, we use another form of Eq.~\eqref{CC_rhodot}, that is
\begin{align}
  \dot{\rho}^2 = \mathrm{e}^{2 (\Phi-\Lambda)} \tilde{L}^2  \left( - \frac{1}{\tilde{L}^2} + \frac{\mathrm{e}^{-2 \Phi}}{b^2} - \frac{\mathrm{e}^{2 \Phi}}{\rho^2} \right) , \label{cc_rhodot2}
\end{align}
where $b := \tilde{L} / \tilde{E}$ is the impact parameter. Since $\dot{\rho}=0$ at the periapsis $\rho = \rho_p$ and the apoapsis $\rho = \rho_a$, we can get two equations from Eq.~\eqref{cc_rhodot2},
\begin{align}
  0 = - \frac{1}{\tilde{L}^2} + \frac{\mathrm{e}^{-2 \Phi_p}}{b^2} - \frac{\mathrm{e}^{2 \Phi_p}}{\rho_p^2} , \quad
  0 = - \frac{1}{\tilde{L}^2} + \frac{\mathrm{e}^{-2 \Phi_a}}{b^2} - \frac{\mathrm{e}^{2 \Phi_a}}{\rho_a^2} ,
\end{align}
where $\Phi_p :=  - M/\rho_p$ and $\Phi_a := -M/\rho_a$. Solving these equations for $b^{2}$ and $\tilde{L}^{2}$, we obtain
\begin{align}
 b^2 = \frac{ \mathrm{e}^{-2 \Phi_p} - \mathrm{e}^{-2 \Phi_a}}{ \mathrm{e}^{2 \Phi_p} / \rho_p^2 - \mathrm{e}^{2 \Phi_a} / \rho_a^2 } , \quad
 \tilde{L}^2 = \frac{\mathrm{e}^{2 \Phi_a} -\mathrm{e}^{2 \Phi_p}}{ \mathrm{e}^{4 \Phi_p}/\rho_p^2 -\mathrm{e}^{4 \Phi_a}/\rho_a^2}.
\end{align}
From Eqs.~\eqref{CC_phidot} and \eqref{cc_rhodot2}, we can obtain the change in $\phi$ as the particle moves from the periapsis to the next periapsis as
\begin{align}
  \delta \phi_\text{C} := 2 \int_{\rho_p}^{\rho_a} \left| \dv{\phi}{\rho} \right| \dd{\rho} = 2 \int_{\rho_p}^{\rho_a} \frac{\dd{\rho}}{\rho^2 \1 \mathrm{e}^{\frac{1}{2} M^2/\rho^2 } \sqrt{- \frac{1}{\rho^2} - \frac{1}{\tilde{L}^2} \mathrm{e}^{2 M/ \rho } + \frac{1}{b^2} \mathrm{e}^{4 M/ \rho }   } }.
  \label{cc_kinseki_0624}
\end{align}
Then, the periapsis shift per round is defined by subtracting the contribution of Newtonian gravity $2 \pi$ from $\delta \phi_\text{C}$:
\begin{align}
  \Delta \phi_\text{C} := \delta \phi_\text{C} - 2 \pi
\end{align}

Here, let us define the eccentricity, the semi-major axis, and the semi-latus rectum as in Eq.~\eqref{kidoupara_teigi1} but in terms of the different radial coordinate $\rho$:
\begin{align}
  e := \frac{\rho_a - \rho_p }{ \rho_a + \rho_p} , \quad  d := \frac{\rho_p + \rho_a}{2} , \quad p := \frac{2 \1 \rho_p \1 \rho_a}{\rho_p + \rho_a} = d(1-e^2) . \label{kidoupara_teigi_cc}
\end{align}
Then $\rho_p$ and $\rho_a$ are written as
\begin{align}
 \rho_p = d(1-e) = p/(1+e) , \quad \rho_a = d(1+e) = p/(1-e). \label{kidoupara_teigi_cc2}
\end{align}
Considering the variable transformation $1/\rho = (1 + e \sin \chi)/p$, Eq.~\eqref{cc_kinseki_0624} becomes
\begin{align}
  \delta \phi_\text{C} = 2 \int_{-\pi/2}^{\pi/2} F(\chi) \dd{\chi} , \label{cc_deltaphi1}
\end{align}
where we have defined the integrand as
\begin{align}
    F(\chi) := \frac{(e/p) \cos \chi }{\mathrm{e}^{\frac{1}{2} M^2 (1+e \sin \chi)^2/p^2 } \sqrt{- (1+e \sin \chi)^2/p^2 - \frac{1}{\tilde{L}^2} \mathrm{e}^{2 M (1+e \sin \chi)/p } + \frac{1}{b^2} \mathrm{e}^{4 M (1+e \sin \chi)/p }   } } . \label{cc_integrand1}
\end{align}

Second, let us implement the step~(II). Expanding $F(\chi)$ in powers of the eccentricity $e$, we obtain
\begin{align}
  F(\chi) = \frac{\mathrm{e}^{-\frac{1}{2}(M/p)^2}}{\sqrt{1-6M/p + 4 (M/p)^2}} + \mathcal{O}(e).
\end{align}
Factoring out the zeroth term in this equation, Eq.~\eqref{cc_deltaphi1} is rewritten as
\begin{align}
  \delta \phi_\text{C} = \frac{2 \1 \mathrm{e}^{-\frac{1}{2}(M/p)^2}}{\sqrt{1-6M/p + 4 (M/p)^2}} \int_{-\pi/2}^{\pi/2}  G(\chi) \dd{\chi} , \label{cc_deltaphi2}
\end{align}
where
\begin{align}
    G(\chi) := \frac{\sqrt{1-6M/p + 4 (M/p)^2}}{\mathrm{e}^{-\frac{1}{2}(M/p)^2}} F(\chi) .
\end{align}
Then, expanding $G(\chi)$ in powers of $e$, we can find 
\begin{align}
  G(\chi) &= G_0(\chi) + G_1(\chi) \1 \frac{M/p}{1-6M/p + 4 (M/p)^2} \2 e + G_2(\chi) \1 \frac{M^2/p^2}{\left\{1-6M/p + 4 (M/p)^2 \right\}^2} \2 e^2 + \cdots . \label{cc_g_tenkai}
\end{align}
The coefficients $G_n(\chi)$ ($n=0$, $1$, $2$, $3$, $\cdots$) are given by
\begin{align}
  G_0(\chi) &:=1, \quad G_1(\chi) := \frac{1}{3}  \left[ 11 + 6 M/p -12 (M/p)^2 \right] (M/p) \sin \chi , \\[8pt]
  G_2(\chi) &:= -\frac{1}{3} \left[ 6 -45 M/p + 80 \1 (M/p)^2 - 48 \1 (M/p)^3 + 8 \1 (M/p)^4 \right] \nonumber \\
  &\hspace{20truemm} + \frac{1}{6} \bigl[ -3 + 66 M/p -169 (M/p)^2 + 252 (M/p)^3 \nonumber \\
  &\hspace{40truemm} -140 (M/p)^4 -48 (M/p)^5 + 48 (M/p)^6  \bigr] \sin ^2 \chi ,
\end{align}
and so forth.

Third, we implement the step~(III). Focusing on each term in Eq.~\eqref{cc_g_tenkai}, we define the new expansion parameter as
\begin{align}
  \alpha := \frac{e \1 M/p}{1 - 6M/p + 4 (M/p)^2} . \label{CC_new_para}
\end{align}
It is noteworthy to point out that, in the case of $e=0$, the non-dimensional quantity in the denominator of $\alpha$ is exactly equal to the left-hand side in the inequality~\eqref{cc_eq_isco} which determines the radius of the ISCO. This definition of the expansion parameter is analogous to the one in Walters' work. Eq.~\eqref{cc_deltaphi2} can be expanded in powers of $\alpha$ as
\begin{align}
  \delta \phi_\text{C} \simeq \frac{2 \1 \mathrm{e}^{-\frac{1}{2}(M/p)^2}}{\sqrt{1 - 6M/p + 4 (M/p)^2}}  \sum_{n=0}^{N-1}  \alpha^{n} \int_{-\pi/2}^{\pi/2} \dd{\chi} G_{n}(\chi). \quad (N=1,2,3,\cdots) \label{cc_kin1}
\end{align}
Although it is difficult to find the general expansion coefficients $G_n(\chi)$, in principle the integrand can be expanded to any order. Whether the expression of general coefficients is easy to obtain or not depends on the spacetime. 

Finally, let us integrate Eq.~\eqref{cc_kin1} and obtain a new series representation for the periapsis shift (step~(IV)). Note that we can find that $G_n$ with odd $n$ up to at least the ninth order (i.e., $G_1, G_3, \cdots, G_9$) are polynomial by odd orders of $\sin \chi$. Therefore these coefficients do not contribute to the integral. Then Eq.\eqref{cc_kin1} can be integrated as
\begin{align}
  \delta \phi_\text{C} \simeq \frac{2 \pi \1  \mathrm{e}^{- \frac{1}{2} (M/p)^2}}{\sqrt{1 - 6 M/p + 4 (M/p)^2}} \sum_{n=0}^{N-1} C_{2n} \1  \alpha^{2n},
\end{align}
where
\begin{align}
  C_0 &:= 1 , \quad C_2 := \frac{1}{12} \Bigl[-27 + 246 M / p - 489 (M/p)^2 + 444 (M/p)^3 \nonumber \\
  &\hspace{50truemm} - 172 (M/p)^4 - 48 (M/p)^5 +48 (M/p)^6 \Bigr] , \\[8pt]
  C_4 &:= \frac{1}{8640} \Bigl[ 30645 -722556 M/p +5231562 (M/p)^2 - 15075756 (M/p)^3 + 23202055 (M/p)^4  \nonumber \\
  &\hspace{10truemm} -21692760 (M/p)^5 + 12242808 (M/p)^6 -2301696 (M/p)^7 - 2423472 (M/p)^8 \nonumber \\
  &\hspace{15truemm} + 1992960 (M/p)^9 - 443520 (M/p)^{10} - 69120 (M/p)^{11} + 34560 (M/p)^{12} \Bigr],
\end{align}
and so forth. The specific forms of the higher order coefficients are omitted here because of their length. It is also possible in principle to obtain any higher order coefficients.

Then we finally obtain a new series representation for the periapsis shift per round in the Chazy-Curzon spacetime as
\begin{align}
  \Delta \phi_\text{C} (N) = 2 \pi \left[\frac{\mathrm{e}^{- \frac{1}{2} (M/p)^2}}{\sqrt{1 - 6 M/p + 4 (M/p)^2}} \sum_{n=0}^{N-1} C_{2n} \1  \alpha^{2n}  -1 \right]. \label{cc_kin_f}
\end{align}

Expanding Eq.~\eqref{cc_kin_f} in powers of $M/p$, we obtain the PN expansion formula~\cite{Bini:2005dy,Vogt:2008zs} (see Appendix~\ref{ap:kin_higher}). Furthermore, in the case of $e=0$, Eq.~\eqref{cc_kin_f} gives an exact formula for the periapsis shift of the quasi-circular orbit:
\begin{align}
  \Delta \phi_\text{C,qc} = 2 \pi \left[\frac{\mathrm{e}^{- \frac{1}{2} (M/d)^2}}{\sqrt{1 - 6 M/d + 4 (M/d)^2}} - 1 \right]. \label{cc_kin_f_qc}
\end{align}
As in the case of the Schwarzschild and the Kerr spacetimes, this formula diverges in the limit of the ISCO since the quantity in the denominator of the first term is exactly equal to the left-hand side in the inequality~\eqref{cc_eq_isco}. 

\subsubsection{Behavior of the expansion parameter} \label{subsec:behave_cc_para}
\begin{figure}[htbp]
  \begin{minipage}{0.45\linewidth}
    \centering
    \includegraphics[keepaspectratio, scale=0.45]{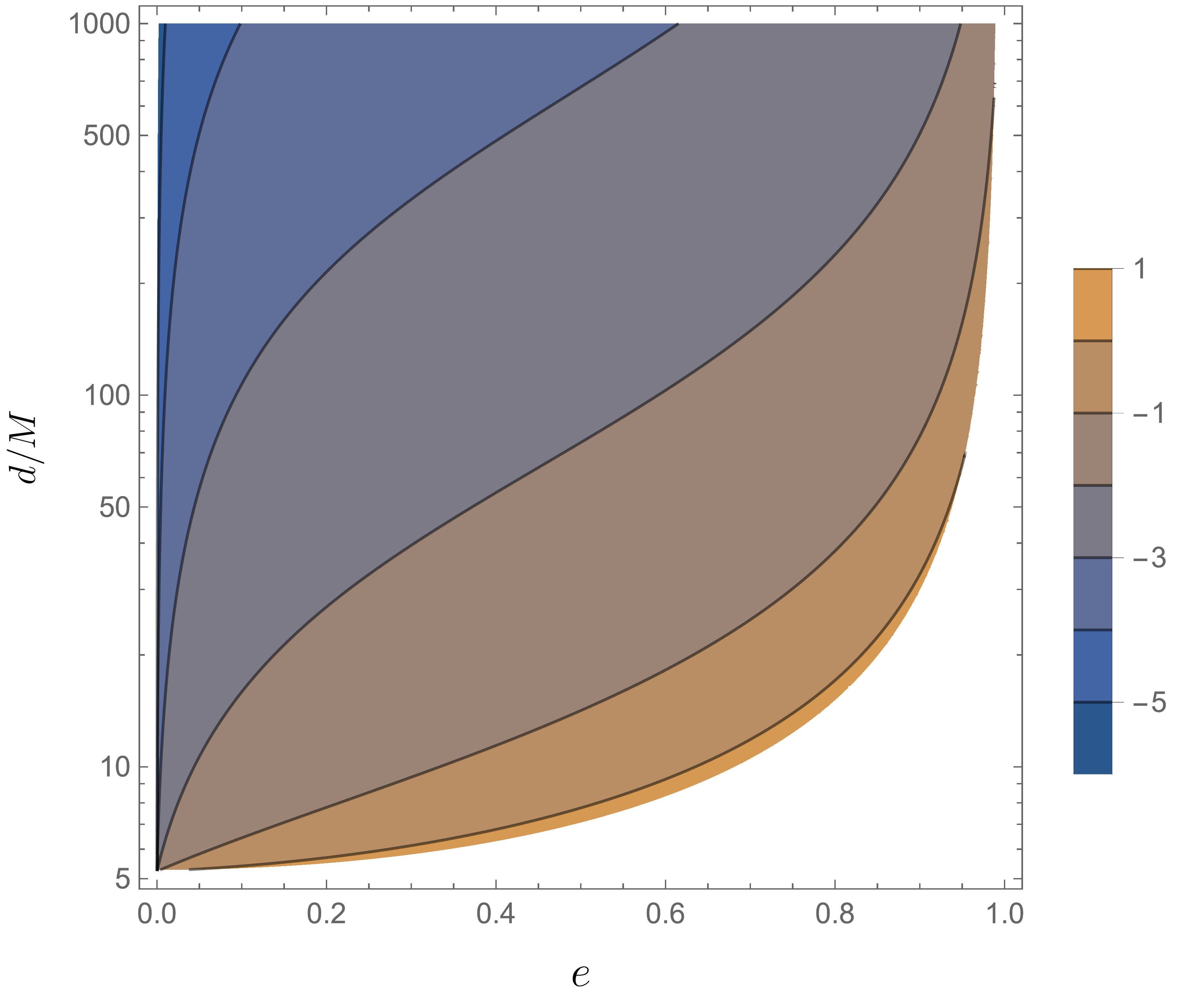}
    \subcaption{New expansion parameter}
    \label{fig:para_CC_para_contour_a}
  \end{minipage} 
  \hspace{8truemm}
  \begin{minipage}{0.45\linewidth}
    \centering
    \includegraphics[keepaspectratio, scale=0.45]{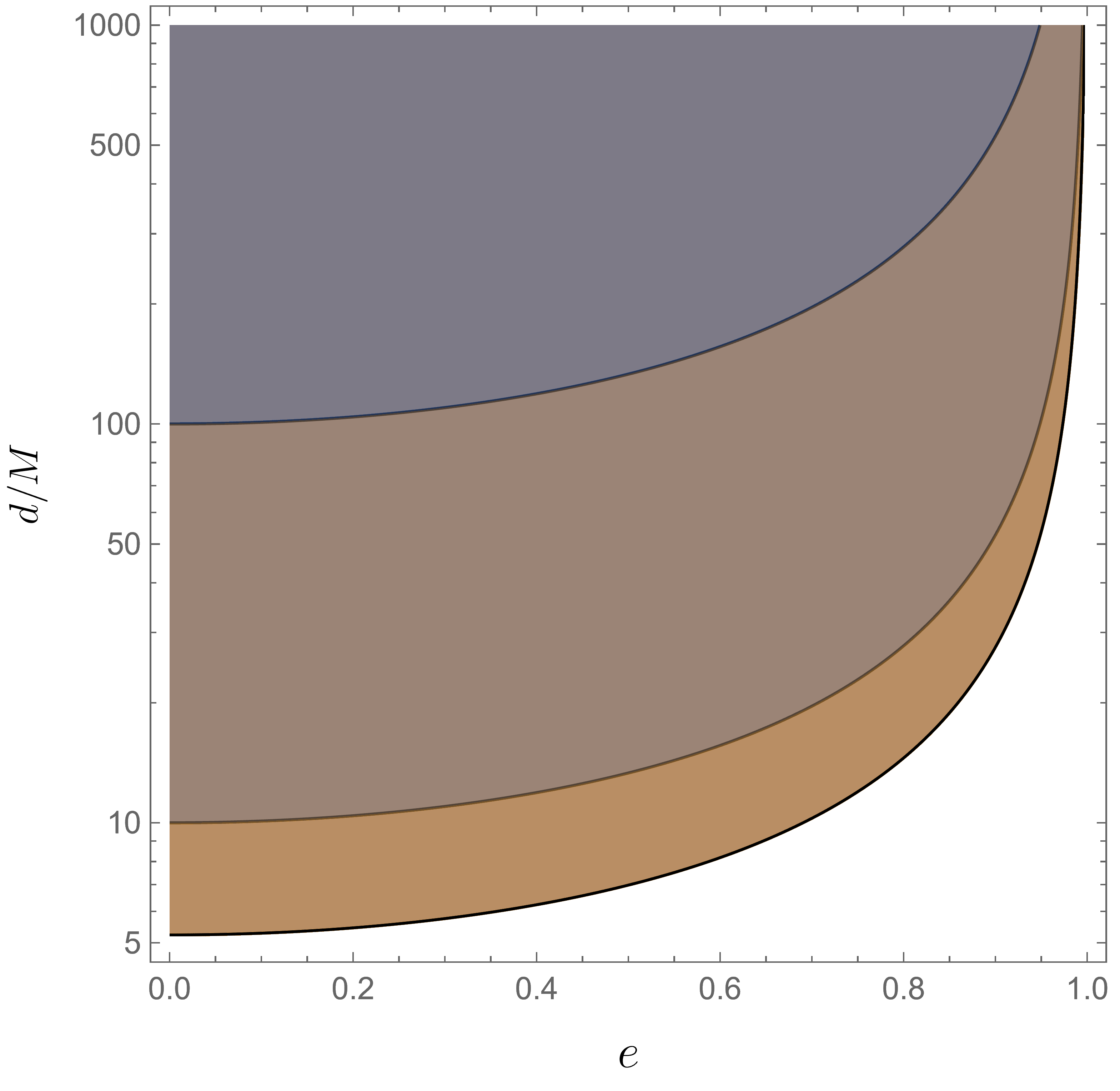}
    \subcaption{PN expansion parameter}
    \label{fig:para_CC_para_contour_b}
  \end{minipage}
  \caption{Contour plots of the new expansion parameter $\alpha$ and the PN expansion parameter $M/p=M/[d(1-e^2)]$. Base 10 logarithms of the parameters are plotted. In the case of $e \to 0$, $\alpha$ approaches zero, while the PN expansion parameter approaches a non-zero value. It can be seen that the larger $d/M$ also leads to the smaller both parameters. Note that the parameter regions that do not satisfy the conditions $1-6M/p+4(M/p)^2>0$, $\rho_a > \rho_\text{ISCO}$, and $L^2>0$, are filled in white.}
  \label{fig:para_CC_contour}
\end{figure}

Here, we discuss the behavior of the new expansion parameter $\alpha$. We show the contour plots of $\alpha$ and the PN expansion parameter $M/p=M/[d(1-e^2)]$ in Fig.~\ref{fig:para_CC_contour}. The main features of the dependence are similar to those of the Kerr case (see Fig.~\ref{fig:para_kerr_para_contour}). We can see that both $\alpha$ and the PN expansion parameter decrease as $e$ is decreased. In the case of $e \to 0$, the $\alpha$ approaches zero since it is proportional to $e$ (see Eq.~\eqref{CC_new_para}), while the PN expansion parameter approaches a non-zero value $M/d$. It can also be seen that the larger $d/M$ leads to the smaller $\alpha$ and the PN expansion parameter.

\subsubsection{Accuracy of the truncated new representation}
Let us investigate the dependence of the accuracy of the formula that is obtained by truncating the new series representation up to a finite number of terms, and compare it with that of the truncated PN expansion formula. Now, we define the periapsis shift per round as the integral form:
\begin{align}
  I := 2 \int_{-\pi/2}^{\pi/2} F(\chi) \dd{\chi} - 2 \pi .
\end{align}
In addition, let us summarize the PN expansion formula~\eqref{cc_kin_pn_higher} as the series in powers of $M/p$:
\begin{align}
  \Delta \phi_\text{C,PN} (N) := \sum_{n=1}^{N} \mathcal{D}_{n}, \label{cc_ps_kyuusuu}
\end{align}
where the expansion coefficients $\mathcal{D}_n$ are given by
\begin{align}
  \mathcal{D}_1 := \frac{6 \1 \pi M}{p} , \quad \mathcal{D}_2 := \frac{\pi \left(44 - 9 \1 e^2\right) M^2}{2 \1 p^2} , \quad \mathcal{D}_3 := \frac{\pi \left( 96 - 53 \1 e^2/2 \right) M^3}{p^3} ,
\end{align}
and so forth. See Eq.~\eqref{cc_kin_pn_higher} in Appendix.~\ref{ap:kin_higher} for the further higher order terms. Eq.~\eqref{cc_ps_kyuusuu} represents the PN expansion formulae taken up to the $N$-th order of $M/p$. Furthermore, we define the relative errors for the truncated new series representation and the truncated PN expansion formula as
\begin{align}
  \Delta_\text{C,NR}(N) := \biggl| \2 \frac{\Delta \phi_\text{C} (N)}{I} -1 \2 \biggr| , \quad 
  \Delta_\text{C,PN} (N) := \biggl| \2 \frac{\Delta \phi_\text{C,PN} (N) }{I} -1 \2 \biggr| ,
\end{align}
respectively.

\begin{figure}[htbp]
\hspace{-10truemm}
  \begin{minipage}{0.45\linewidth}
    \centering
    \includegraphics[keepaspectratio, scale=1.09]{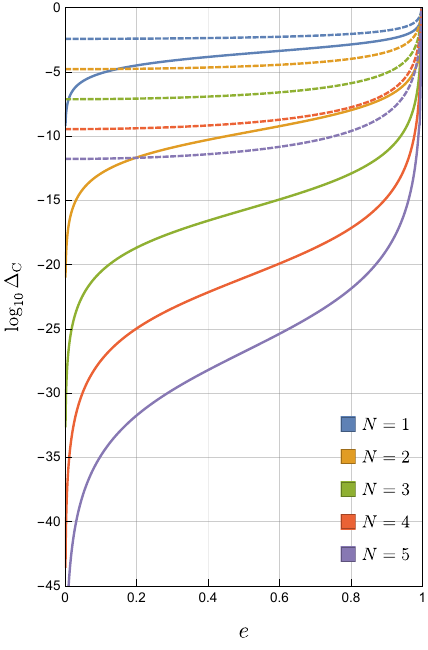}
    \subcaption{$d/M=1000$}
    \label{fig:CCgosaeplotd1000}
  \end{minipage} 
  \hspace{7truemm}
  \begin{minipage}{0.45\linewidth}
    \centering
    \includegraphics[keepaspectratio, scale=1.09]{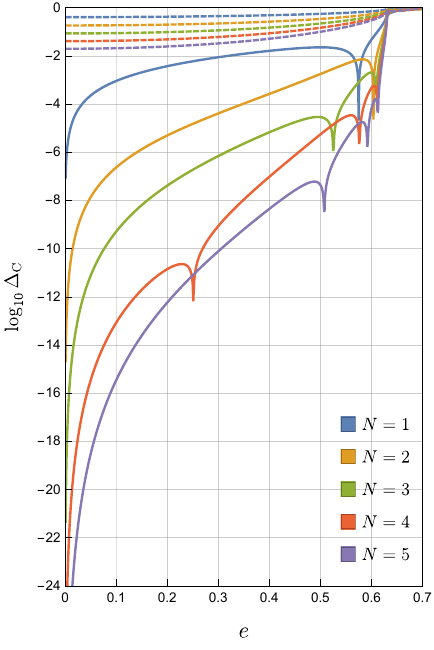}
    \subcaption{$d/M=10$}
    \label{fig:CCgosaeplotd10}
  \end{minipage}
  \caption{Dependence of the relative errors of the truncated new representation $\Delta_\text{C,NR}(N)$ and the truncated PN expansion formula $\Delta_\text{C,PN}(N)$ on the eccentricity $e$. Solid and dashed lines represent the errors $\Delta_\text{C,NR}(N)$ and $\Delta_\text{C,PN}(N)$, respectively. It can be seen that in the case of $e \to 0$, the errors of the truncated new representation approach zero, while the errors of the truncated PN expansion formula approach to non-zero values.}
  \label{fig:CCgosaeplot}
\end{figure}

\begin{figure}[htbp]
\hspace{-10truemm}
  \begin{minipage}{0.45\linewidth}
    \centering
    \includegraphics[keepaspectratio, scale=1.09]{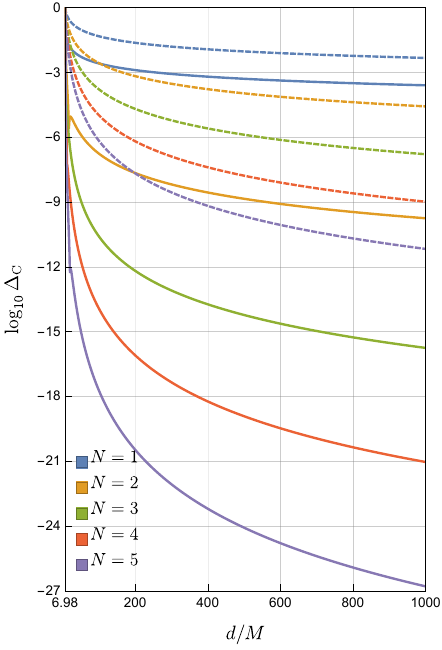}
    \subcaption{$e=1/2$}
    \label{fig:CCgosadplote05}
  \end{minipage} 
  \hspace{7truemm}
  \begin{minipage}{0.45\linewidth}
    \centering
    \includegraphics[keepaspectratio, scale=1.09]{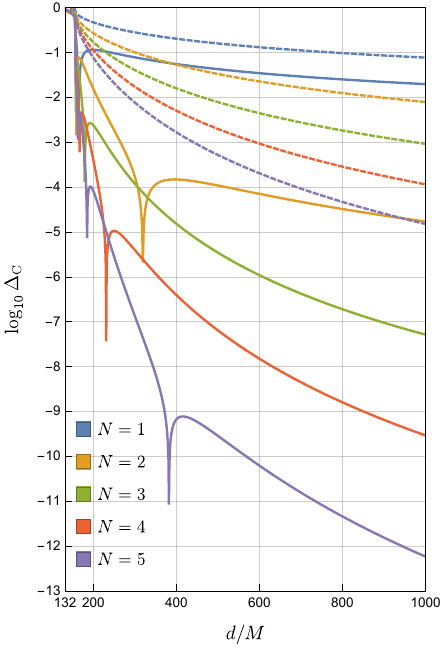}
    \subcaption{$e=0.98$}
    \label{fig:CCgosadplote098}
  \end{minipage}
  \caption{Dependence of the relative errors of the truncated new representation $\Delta_\text{C,NR}(N)$ and the truncated PN expansion formula $\Delta_\text{C,PN}(N)$ on the semi-major axis $d$. Solid and dashed lines represent the errors $\Delta_\text{C,NR}(N)$ and $\Delta_\text{C,PN}(N)$, respectively. It can be seen that the error of the truncated new representation decreases as $d/M$ is increased. The origins of the horizontal axis are determined by using the condition $1-6M/p+4(M/p)^2>0$.}
  \label{fig:CCgosadplot}
\end{figure}

Figs.~\ref{fig:CCgosaeplot} and \ref{fig:CCgosadplot} show the dependence of the relative errors of the truncated new representation $\Delta_\text{C}(N)$ and the truncated PN expansion formula $\Delta_\text{C,PN}(N)$ on the eccentricity $e$ and the semi-major axis $d$, respectively. Roughly speaking, the dependence of the error of the truncated new representation is similar to the Kerr case (see Figs.~\ref{fig:Kerrgosaeplot} and~\ref{fig:Kerrgosadplot}). First, let us focus on the dependence of the error on $e$. From Fig.~\ref{fig:CCgosaeplotd1000}, we can see that the error of the truncated new representation (solid lines) monotonically decreases with decreasing $e$. In addition, the larger $N$ also leads to the higher accuracy. Note that the accuracy rapidly worsens when $e$ approaches 1. Dashed lines in Fig.~\ref{fig:CCgosaeplotd1000} represent the errors of the truncated PN expansion formula. We can see that the error of the truncated PN expansion formula also decreases as $e$ is decreased. The major difference in the dependence of the accuracy of the truncated new representation and the truncated PN expansion formula on $e$ appears in the case of $e \to 0$. In this case, the error of the truncated new representation rapidly approaches zero, while the error of the truncated PN expansion formula approaches a non-zero value. This difference would be caused by the dependence of the expansion parameters on $e$ since we have found a similar difference in the behavior of the expansion parameters in Sec.~\ref{subsec:behave_cc_para}. In addition, Fig.~\ref{fig:CCgosaeplotd10} shows the dependence of the errors on $e$ in the case where the orbit is closer to the center ($d/M=10$). It can be seen that even if the PN expansion formula is taken up to $N=5$ and $e$ is close to 0, the relative error is reduced to only $10^{-2}$ at most approximately. On the other hand, the relative error of the truncated new representation is less than approximately $10^{-7}$ in the case of $e \to 0$, even if $N=1$. We can also find that as $N$ is increased, the accuracy of the truncated new representation is maintained even if $e$ is increased to some extent. From Fig.~\ref{fig:CCgosaeplotd10}, we can find that if $e$ is sufficiently small, the truncated new representation has higher accuracy than the truncated PN expansion formula even if the orbit is close to the center where the convergence of the PN expansion formula is not guaranteed.

Next, we discuss the dependence of the accuracy on the semi-major axis $d$. From Fig.~\ref{fig:CCgosadplote05}, we can see that the larger $d/M$ and larger $N$ lead to the smaller error of the truncated new representation (solid lines), respectively. Dashed lines in Fig.~\ref{fig:CCgosadplote05} represent the errors of the truncated PN expansion formula for comparison. It can be seen that the error of the truncated PN expansion formula also has a similar behavior to the truncated new representation. Here, we emphasize that the truncated new representation has higher accuracy than the truncated PN expansion formula if we compare for the same $N$ in Fig.~\ref{fig:CCgosadplote05}. In addition, Fig.~\ref{fig:CCgosadplote098} shows the dependence of the errors on $d/M$ in the case of the orbit is highly eccentric ($e=0.98$). It is noteworthy to mention that even if the orbit is highly eccentric, the truncated new representation has higher accuracy than the truncated PN expansion formula if $d/M$ and $N$ are sufficiently large. This is also one of the advantages of the truncated new representation. 

\subsection{Application to the Kerr spacetime}
Here, we discuss the application of the present method to the Kerr spacetime. In Sec.~\ref{sec:peri_kerr}, for the case of the Kerr spacetime, we have defined the expansion parameter $\gamma$ as Eq.~\eqref{gamma_def} motivated by Walters' work. It is noteworthy to point out that this parameter $\gamma$ is proportional to $e$ if we fix $p$ and $r_0$ as seen in the middle expression of Eq.~\eqref{gamma_def}. That is to say, the parameter $\gamma$ can be naturally found by expanding in $e$ with fixed $p/r_0$ at step~(III) in the general method. This means that the new series representation for the Kerr spacetime obtained in Sec.~\ref{sec:peri_kerr} can be obtained by the general method. In this sense, the new series expansion of the periapsis shift in the Kerr spacetime which we have proposed in Sec.~\ref{sec:peri_kerr} is actually a specific application of the general method. On the other hand, if we fix $p/M$, $\gamma$ contains higher orders of $e$. This implies that the expansion parameter is not uniquely determined but has some ambiguity. 

\section{Summary and conclusions}\label{sec:sum}
In this paper, we have proposed a new series expansion method for the periapsis shift by generalizing Walters' work. The method formulates the periapsis shift in various spacetimes as series analytically without using special functions and provides simple and highly accurate approximate formulae by truncating the new series representation up to a finite number of terms.

In Sec.~\ref{sec:peri_kerr}, we have presented a new series expansion of the periapsis shift in the Kerr spacetime. We have defined the new expansion parameter $\gamma$ as Eq.~\eqref{gamma_def} motivated by Walters' work. Using this expansion parameter $\gamma$, we have obtained a new series representation for the periapsis shift in the Kerr spacetime. The new series representation reproduces the PN expansion formula, an exact formula for the quasi-circular orbit, and Walters' representation. The dependence of $\gamma$ on the orbital parameters was also discussed. The $\gamma$ monotonically decreases with decreasing $e$. The larger semi-major axis per mass $d/M$ also leads to the smaller $\gamma$. The major difference in the dependence on $e$ between $\gamma$ and the PN expansion parameter $M/p=M/[d(1-e^2)]$ appears in the case of $e \to 0$. In this case, the $\gamma$ approaches zero, while the PN expansion parameter approaches a non-zero value. In addition, we have found that the $\gamma$ monotonically decreases with increasing the non-dimensional Kerr parameter $a/M$ in the prograde case. In contrast, $\gamma$ monotonically decreases with decreasing $a/M$ in the retrograde case. We have also investigated the relative error of the formula that is obtained by truncating the new series representation up to a finite number of terms. We have found that the error of the truncated new representation decreases as $e$ is decreased. The larger $d/M$ also leads to the higher accuracy. Additionally, in the prograde case, the larger $a/M$ leads to the higher accuracy. In contrast, in the retrograde case, the larger $a/M$ leads to the lower accuracy. The major difference in the behavior of the errors between the truncated new representation and the truncated PN expansion formula is apparent in the case of $e \to 0$. In this case, the relative error of the truncated new representation approaches zero, while the relative error of the truncated PN expansion formula approaches non-zero values. This dependence on $e$ is similar to that of $\gamma$.

In Sec.~\ref{sec:new_method}, we have proposed a general method for the new series expansion of the periapsis shift in various spacetimes by generalizing Walters' work. In the new method, the expansion parameter is defined as the eccentricity divided by the quantity that vanishes in the limit of the ISCO. This definition of the expansion parameter is analogous to the one in Walters' work. We have confirmed that the new series expansion of the periapsis shift in the Kerr spacetime which we have discussed in Sec.~\ref{sec:peri_kerr} is actually a specific application of the general method. As another specific example, we have considered the application of the general method to the Chazy-Curzon spacetime and obtained a new series representation for the periapsis shift. An exact formula for the periapsis shift of the quasi-circular orbit has also been obtained by taking the limit of the circular orbit. We have examined the behavior of the expansion parameter $\alpha$ and the relative error of the truncated new representation. Similar to the expansion parameter $\gamma$ in the Kerr case, $\alpha$ decreases as $e$ is decreased. The larger $d/M$ also leads to the smaller $\alpha$. The relative error of the truncated new representation also has a similar dependence as $\alpha$. That is to say, the relative error approaches zero in the case of $e \to 0$, while the error of the truncated PN expansion formula approaches a non-zero value. The larger $d/M$ also leads to the smaller relative error of the truncated new representation. 

As mentioned above, the new series expansion method formulates the periapsis shift in various spacetimes as series analytically without using special functions and provides simple and highly accurate approximate formulae by truncating the new series representations up to a finite number of terms. In the method, the expansion parameter is defined as the eccentricity divided by the non-dimensional quantity that vanishes in the limit of the ISCO. This means that the expansion parameter denotes how eccentric the orbit is and how close it is to the ISCO. That's why the smaller the eccentricity, the higher the accuracy of the truncated new representation. If the eccentricity is sufficiently small, the truncated new representation has higher accuracy than the truncated PN expansion formula even in strong gravitational fields where the convergence of the PN expansion formula is not guaranteed. In addition, even if the orbit is highly eccentric, the truncated new representation has comparable or higher accuracy than the truncated PN expansion formulae if the semi-major axis per mass is sufficiently large. The highly accurate, simple, and easy-to-use formulae obtained by truncating the new series representations up to a finite number of terms are useful, for example, for discussing and calculating the periapsis shifts in strong gravitational fields and checking a numerical integration. The above are the merits of the new series expansion method and the new series representations obtained by it. Applications of the method to other spacetimes such as the Kerr-Newman spacetime are to be investigated.

\acknowledgments
The authors are grateful to Y. Hatsuda, T. Hiramatsu, T. Igata, T. Ishii, T. Kobayashi, J. Saito, and R. Takahashi for their helpful
comments. This work was supported by Rikkyo University Special Fund for Research (A.K.), JSPS KAKENHI Grant Nos.~JP20H05853, JP24K07027 (T.H.) and JP20K14467 (K.O.).
\appendix
%
\renewcommand{\theequation}{A.\arabic{equation}}
\makeatletter
\@addtoreset{equation}{section}
\makeatother
\section{Periapsis shift in the Schwarzschild spacetime}\label{ap:Sch_kin}
\subsection{Walters' representation} \label{ap:walters}
Let us review Walters' work~\cite{walters2018simple}, which derives a new exact series representation for the periapsis shift in the Schwarzschild spacetime~\cite{Sch1916}. The line element in the Schwarzschild spacetime is given by
\begin{align}
  \dd{s}^2 = -\left(1-\frac{2M}{r} \right) \dd{t}^2 + \left(1-\frac{2M}{r} \right)^{-1} \dd{r}^2 + r^2 \dd{\theta}^2 + r^2 \sin^2 \theta \dd{\phi}^2, 
\label{met_sch}
\end{align}
where $M$ denotes the mass of the BH. Now we assume $r>2M$. Let us consider the geodesic motion of a test particle with the mass $m$. By spherical symmetry, we can assume that the orbit is on the equatorial plane (i.e., $\theta = \pi/2$) without loss of generality. The Lagrangian of the particle is given by
\begin{align}
 \mathcal{L} =  \frac{1}{2} \1 m   \Biggl[ -\biggl(1-\frac{2M}{r} \biggr) \1 \dot{t}^2 + \biggl(1-\frac{2M}{r} \biggr)^{-1} \dot{r}^2 +  r^2 \dot{\phi}^2 \Biggr],
  \label{lag_sch}
\end{align}
where the dot denotes the derivative with respect to the proper time. Since the spacetime symmetry, the energy and the angular momentum 
\begin{align}
   E := - \pdv{\mathcal{L}}{\dot{t}} , \quad L := \pdv{\mathcal{L}}{\dot{\phi}} ,
\end{align}
are conserved. Evaluating the derivatives from Eq.~\eqref{lag_sch} and solving for $\dot{t}$ and $\dot{\phi}$, we can get
\begin{align}
  \dot{t} = \frac{\tilde{E}}{1-2M/r} , \quad \dot{\phi} = \frac{\tilde{L}}{r^2} , \label{sdp}
\end{align}
where $\tilde{E} := E / m $ and $\tilde{L} := L / m $. Furthermore, substituting Eq.~\eqref{sdp} into the normalization condition $g_{\mu \nu} \1 \dot{x}^\mu \dot{x}^\nu = -1$, we obtain after some simplification
\begin{align}
  \dot{r}^2 &= \frac{2 M \tilde{L}^2}{r^3} - \frac{\tilde{L}^2}{r^2} + \frac{2M}{r} - (1-\tilde{E}^2) . \label{sch_rdot}
\end{align}
From Eqs.~\eqref{sdp} and \eqref{sch_rdot}, we find
\begin{align}
  \left| \dv{r}{\phi} \right| = r^2 \sqrt{R_\text{S} (r)}, \label{drdphi}
\end{align}
where
\begin{align}
  R_\text{S}(r) := \frac{2M }{r^3} - \frac{1}{r^2} + \frac{2M}{\tilde{L}^2 \1 r} -\frac{1 - \tilde{E}^2}{\tilde{L}^2} . \label{diff_R}
\end{align}
Hereafter, we assume the stable bound orbit and $\tilde{E}^2 < 1$. Then we can find that the function $R_\text{S} (r)$ has three real zeros, which correspond to the turning points of the orbit. Let us express the zeros as $r_0$, $r_p$, and $r_a$ in order from smallest to largest (i.e., $r_0 < r_p < r_a$). Note that the $r_p$ and $r_a$ denote the periapsis and apoapsis, respectively, and the bound orbit exists between these turning points. The function $R_\text{S} (r)$ can be factorized as
\begin{align}
  R_\text{S} (r) = 2 M \left( \frac{1}{r_0}- \frac{1}{r} \right) \left( \frac{1}{r_p}- \frac{1}{r} \right) \left( \frac{1}{r}- \frac{1}{r_a} \right) . \label{sch_Rin}
\end{align}
Comparing the coefficients of Eqs.~\eqref{diff_R} and \eqref{sch_Rin}, we obtain three equations:
\begin{align}
  1 = 2 M \left(\frac{1}{r_0}+\frac{ r_p + r_a}{r_p \1 r_a} \right) , \quad \frac{1}{\tilde{L}^2} = \frac{r_p+r_a+r_0}{r_p \1 r_a \1 r_0  } , \quad \frac{1-\tilde{E}^2}{\tilde{L}^2} = \frac{2 M}{ r_p \1 r_a \1 r_0 } . \label{Sch_renritu}
\end{align}
Note that we can find $r_0 > 0$ from the third equation. Solving these equations for $\tilde{E}$, $\tilde{L}$, and $r_0$, we obtain
\begin{align}
  \tilde{E} &= \sqrt{ \frac{(r_p + r_a)(r_p - 2M)(r_a - 2M)}{r_p \1 r_a (r_p + r_a) - 2 M (r_p^2 + r_p \1 r_a + r_a^2 ) } } , \label{sch_E} \\[8pt]
  \tilde{L} &= \pm \sqrt{ \frac{2M r_p^2 \1 r_a^2}{r_p \1 r_a (r_p + r_a) - 2 M (r_p^2 + r_p \1 r_a + r_a^2 ) } } , \label{sch_L} \\[8pt]
  r_0 &= \frac{2M r_p \1 r_a}{r_p \1 r_a - 2M(r_p + r_a)} . \label{sch_r0}
\end{align}
Note that these formulae are recovered from Eqs.~\eqref{tilEne}, \eqref{tilL}, and~\eqref{r0proret} in the case of $a=0$. Here, let us define the eccentricity $e$, the semi-major axis $d$, and the semi-latus rectum $p$ by using $r_p$ and $r_a$ as in Eq.~\eqref{kidoupara_teigi1}. By using these orbital parameters, Eqs.~\eqref{sch_E}, \eqref{sch_L}, and \eqref{sch_r0} can be rewritten as
\begin{align}
  \tilde{E} = \sqrt{ \frac{1 - 4 \left\{1 - (1-e^2) M/p \right\} M/p }{1 -(3+e^2) M /p } } , \quad 
  \tilde{L} = \pm \sqrt{ \frac{ M p }{1 -(3+e^2) M / p} } , \quad
  r_0 = \frac{2 M }{1-4 M/p} . \label{sch_r0_f2}
\end{align}
From Eqs.~\eqref{drdphi} and \eqref{sch_Rin}, we have
\begin{align}
  \left| \dv{\phi}{r} \right| = \frac{1}{r^2 \sqrt{ 2 M \bigl( \frac{1}{r_0}- \frac{1}{r} \bigr) \bigl( \frac{1}{r_p}- \frac{1}{r} \bigr) \bigl( \frac{1}{r}- \frac{1}{r_a} \bigr)  } } . \label{gyakusuu}
\end{align}
A change in $\phi$ as the particle moves from the periapsis to the next periapsis is defined as
\begin{align}
  \delta \phi_\text{S} := 2 \int_{r_p}^{r_a} \left| \dv{\phi}{r} \right| \dd{r} . \label{delta_phi_teigi}
\end{align}
Then, the periapsis shift per round is defined by subtracting the contribution of Newtonian gravity $2 \pi$ from $\delta \phi_\text{S}$:
\begin{align}
  \Delta \phi_\text{S} &:= \delta \phi_\text{S} - 2 \pi. \label{Delta_phi_teigi}
\end{align}
Substituting Eq.~\eqref{gyakusuu} to Eq.~\eqref{delta_phi_teigi}, and introducing $u := 1/r$, we obtain
\begin{align}
  \delta \phi_\text{S} = \frac{2}{\sqrt{2M}} \int_{u_a}^{u_p} \frac{\dd{u}}{ \sqrt{  1/r_0 - u  }  \sqrt{(u_p-u)(u-u_a)}   } , \label{sekibun0}
\end{align}
where $u_p := 1/r_p$ and $u_a := 1/r_a$. Furthermore, the variable transformation $u = (1 + e \sin \chi)/p$ implies 
\begin{align}
  \delta \phi_\text{S} = \frac{2}{\sqrt{ 2M  \left( 1/r_0 - 1/p \right) }} \int_{-\pi/2}^{\pi/2}  \frac{\dd{\chi}}{ \sqrt{  1 - \frac{e}{p/r_0 - 1} \sin \chi  } } . \label{sekibun1}
\end{align}
Here, we define the expansion parameter $\beta$ as the coefficient of $\sin \chi$ in the integrand:
\begin{align}
  \beta := \frac{e}{p/r_0 - 1} = \frac{2 \1 e M / p}{1 - 6 M /p} \label{def_beta} .
\end{align}
It is noteworthy that the parameter $\beta$ is defined as the eccentricity $e$ divided by the non-dimensional quantity that vanishes in the limit of the ISCO (i.e., $e = 0$ and $d \to 6M$). This means that $\beta$ represents how eccentric the orbit is and how close it is to the ISCO. 

Rewriting Eq.~\eqref{sekibun1} with $\beta$, we obtain
\begin{align}
  \delta \phi_\text{S} = \frac{2}{\sqrt{1-6M/p}} \, \int_{-\pi/2}^{\pi/2} \frac{\dd{\chi}}{ \sqrt{1 - \beta \sin \chi } } . \label{sch_sekibun3}
\end{align}
By using the expansion formula~\eqref{tenkai_kousiki2}, this becomes
\begin{align}
  \delta \phi_\text{S} = \frac{2}{\sqrt{1-6M/p}} \sum_{n=0}^\infty A_{2n} \1 \beta^{2n} \int_{-\pi/2}^{\pi/2} \dd{\chi} \2 \sin^{2n} \chi . \label{sch_sekibun4}
\end{align}
Note that odd powers of $\sin \chi$ do not contribute to the integral. By using the integral formula~\eqref{sekibun_kousiki3}, Eq.~\eqref{sch_sekibun4} can be integrated as
\begin{align}
  \delta \phi_\text{S} = \frac{2 \1 \pi}{\sqrt{1-6M/p}} \sum_{n=0}^\infty A_n \1 A_{2n} \1 \beta^{2n} . \label{sch_kin1}
\end{align}
Then, we finally obtain a new series representation for the periapsis shift in the Schwarzschild spacetime as
\begin{align}
  \Delta \phi_\text{S} = 2 \1 \pi \left[ \frac{1}{\sqrt{1-6M/p}} \sum_{n=0}^\infty A_n \1 A_{2n} \1 \beta^{2n}  -1 \right] . \label{sch_kin_revolution}
\end{align}
This is Walters' representation.

Expanding Eq.~\eqref{sch_kin_revolution} in powers of $M/p$, we can obtain the PN expansion formula (see e.g., \cite{Einstein:1916vd,Bini:2005dy,Vogt:2008zs,de2011estimating,poisson_will_2014,Tucker:2018rgy,He:2023joa}). We show the expression with further higher order terms in Appendix~\ref{ap:kin_higher}. In addition, in the case of $e = 0$, Eq.~\eqref{sch_kin_revolution} provides an exact formula for the periapsis shift of the quasi-circular orbit~\cite{Schmidt:2008qi,Schmidt:2011xi,walters2018simple,Tucker:2018rgy,Harada:2022uae}
\begin{align}
  \Delta \phi_\text{\1S,qc} = 2 \1 \pi \Biggl[ \frac{1}{\sqrt{1-6M/d}} - 1 \Biggr] \label{sch_peri_shift_qc} .
\end{align}
Note that this formula diverges in the limit of the ISCO (i.e., $d \to 6M$).

Finally, let us discuss the convergence of Walters' representation~\eqref{sch_kin_revolution}. The convergence radius of the series can be calculated as
\begin{align}
  r_\text{c} = \lim_{n \to \infty }\, \Biggl| \frac{A_n \1 A_{2n}}{A_{n+1} \1 A_{2(n+1)}} \Biggr| = 1 .
\end{align}
That is to say, Walters' representation converges for
\begin{align}
 \beta = \frac{e}{p/(2M)-3} < 1.
\end{align}
In the case of $p/(2M)-3 \geq 1$, that is $p \geq 8M$, the series converges for any value of $e$. On the other hand, in the case of $0< p/(2M)-3 <1$, that is $6M < p < 8M$, it also converges for $e < p/(2M)-3$. Qualitatively, if the orbit is far from a central object, the series converges even if the orbit is highly eccentric. On the other hand, even if the orbit is close to a central object, the series also converges if the eccentricity is sufficiently small. The PN expansion formula for the periapsis shift is the expansion in powers of $M/p$, assuming that it is sufficiently small (i.e., weak gravity field approximations). In strong gravitational fields where the orbit is close to the center, the convergence of the PN expansion formula is not guaranteed. Therefore, it is one of the advantages of Walters' representation that the convergence radius is obtained exactly and the convergence is guaranteed even in strong gravitational fields if the eccentricity is sufficiently small.

\subsection{Exact formula by the elliptic integral} \label{ap:peri_sch_daen}
In Walters' work, the integral~\eqref{sekibun0} has been evaluated by the variable transformation with orbital parameters. It should be pointed out that there is another way to evaluate this integral by using the elliptic integral~\cite{synge_1960}. Let us review the approach. Eq.~\eqref{sekibun0} can be written as
\begin{align}
    \delta \phi_\text{S} &= 2 \sqrt{u_g} \int_{u_a}^{u_p} \frac{ \dd{u}}{ \sqrt{ (u - u_0)(u - u_p)(u - u_a)} } , \label{kindaen}
\end{align}
where $u_g := 1/(2M)$ and $u_0 := 1/r_0 = u_g -u_a -u_p$. This integral can be rewritten as
\begin{align}
  \delta \phi_\text{S} &= \frac{4 \sqrt{u_g}}{\sqrt{u_0 -u_a}} \, K \left( \sqrt{\frac{u_p-u_a}{u_0-u_a}} \, \right) \\[8pt]
  &= \frac{4}{\sqrt{1 - 2 \1 (3-e)M/p}} \, K \left( 2 \, \sqrt{\frac{e \1 M/p}{1 - 2 \1 (3-e)M/p }} \, \right) ,
  \label{sch_kandaenseki1}
\end{align}
where $K(k)$ is the complete elliptic integral of the first kind, which is given by
\begin{align}
  K(k) = \int_0^{\pi/2} \frac{\dd{\chi}}{\sqrt{1-k^2 \sin^2 \chi}} = \frac{\pi}{2}  \sum_{n=0}^\infty A_n^{\2 2} \2 k^{2n}  .
\end{align}
The expansion coefficients $A_n$ are defined by Eq.~\eqref{tenkai_kousiki2}. Eq.~\eqref{sch_kandaenseki1} also can be rewritten in the form of an infinite series:
\begin{align}
  \delta \phi_\text{S} &= \frac{2 \1 \pi}{\sqrt{1 - 2 M\1 (3-e)/p}} \sum_{n=0}^\infty A_n^{\2 2} \2 \bar{\beta}^{2n} , \label{sch_kin_daen_f1} \\[8pt]
  \bar{\beta} &:= \sqrt{\frac{u_p-u_a}{u_0-u_a}} = 2 \, \sqrt{\frac{e \1 M /p}{1-2 M \1 (3-e)/p }} .
\end{align}
Then, the periapsis shift per round is written as
\begin{align}
  \Delta \phi_\text{S} = 2 \pi \left[ \frac{1}{\sqrt{1 - 2 \1 (3-e)M/p}} \sum_{n=0}^\infty  A_n^2 \1 \bar{\beta}^{2n} - 1 \right] . \label{sch_kin_daen_revo1}
\end{align}
This is another exact series representation which is similar to Walters' representation~\eqref{sch_kin_revolution}. In the case of $e = 0$, Eq.~\eqref{sch_kin_daen_revo1} gives the formula for the periapsis shift of the quasi-circular orbit which agrees with Eq.~\eqref{sch_peri_shift_qc}. In addition, by expanding Eq.~\eqref{sch_kin_daen_revo1} in powers of $M/p$, we also obtain the PN expansion formula~\eqref{eq:sch_PN_higher}.

\renewcommand{\theequation}{B.\arabic{equation}}
\makeatletter
\@addtoreset{equation}{section}
\makeatother
\section{Post-Newtonian expansion formulae for the periapsis shift} \label{ap:kin_higher}
In this section, we present the PN expansion formulae for the periapsis shift with higher order terms in the Schwarzschild, the Kerr, and the Chazy-Curzon spacetimes. 
\subsection{Schwarzschild spacetime}
Expanding Eq.~\eqref{sch_kin_revolution} or Eq.~\eqref{sch_kin_daen_revo1} in powers of $M/p$, we obtain the PN expansion formula for the periapsis shift in the Schwarzschild spacetime (see e.g., \cite{Einstein:1916vd,Bini:2005dy,Vogt:2008zs,de2011estimating,poisson_will_2014,Tucker:2018rgy,He:2023joa}):
\begin{align}
 \Delta \phi_\text{S,PN} &= \frac{6\1 \pi M}{p}  +\frac{3 \1 \pi \1  (18+e^2)M^2}{2\1 p^2} +\frac{45 \1 \pi \1 (6+e^2)M^3}{2 \1 p^3}  +\frac{105  \1 \pi  \1 (216+72 \1 e^2+e^4)M^4}{32\1 p^4} \nonumber \\[5pt]
 &\quad + \frac{567 \1 \pi \1 (216 +120 \1 e^2+ 5 \1 e^4)M^5}{32 \1 p^5} + \frac{231 \1 \pi \1 (11664 +9720 \1 e^2 + 810 \1 e^4 +5 \1 e^6 ) M^6}{128 \1 p^6} \nonumber \\[5pt]
 &\quad + \frac{1287 \1 \pi \1 (11664 + 13608 \1 e^2 +1890 \1 e^4 +35 \1 e^6) M^7}{128 \1 p^7} \nonumber \\[5pt]
 &\quad + \frac{6435 \1 \pi \1  (839808+1306368 \1 e^2 + 272160 \1 e^4 +10080 \1 e^6+ 35 \1 e^8 ) M^8}{8192 \1 p^8} \nonumber \\[5pt]
 &\quad + \frac{328185 \1 \pi \1 (93312  + 186624\1  e^2 + 54432 \1 e^4 + 3360 \1 e^6 + 35 \1 e^8) M^9}{8192 \1 p^9} \nonumber \\[5pt]
 &\quad + \frac{415701 \1 \pi \1  (1679616 +4199040 \1 e^2 + 1632960 \1 e^4+151200 \1 e^6 + 3150 \1  e^8+ 7 \1 e^{10}) M^{10}}{32768 \1 p^{10}} \nonumber \\[5pt]
 &\quad + \cdots.
\label{eq:sch_PN_higher}
\end{align}

\subsection{Kerr spacetime}
Expanding Eq.~\eqref{tenhou21-4} in powers of $M/p$ and $a/p$, we obtain the PN expansion formula for the periapsis shift in the Kerr spacetime (see e.g., \cite{boyer_price_1965,Bini:2005dy,Vogt:2008zs,de2011estimating,He:2023joa}):
\begin{align}
  &\Delta \phi_\text{K,PN} =  \frac{6 \1 \pi M}{p} + \frac{3 \1 \pi \1 (18+e^2) M^2}{2 \1 p^2} + \frac{45 \1 \pi \1 (6+e^2) M^3}{2 \1 p^3} + \frac{105 \1 \pi \1 (216+72 \1 e^2 + e^4) M^4}{32 \1 p^4} \nonumber \\[5pt]
  &\quad + \frac{567 \1 \pi \1 (216 +120 \1 e^2+ 5 \1 e^4)M^5}{32 \1 p^5} + \frac{231 \1 \pi \1 (11664 +9720 \1 e^2 + 810 \1 e^4 +5 \1 e^6 ) M^6}{128 \1 p^6} \nonumber \\[5pt]
  &\quad + \frac{1287 \1 \pi \1 (11664 + 13608 \1 e^2 +1890 \1 e^4 +35 \1 e^6) M^7}{128 \1 p^7} \nonumber \\[5pt]
  &\quad \mp \Biggl(  \frac{8 \1 \pi \1 a \1 M^{1/2}}{p^{3/2}} + \frac{72 \1 \pi \1 a \1 M^{3/2}}{p^{5/2}} + \frac{54 \1 \pi \1 (10+e^2) \1 a \1 M^{5/2}}{p^{7/2}} + \frac{6 \1 \pi (630 + 173 e^2) \1 a \3 M^{7/2}}{p^{9/2}} \nonumber \\[5pt]
  &\quad + \frac{3 \1 \pi \left(68040 +34936 e^2 + 739 e^4\right) \1 a \3 M^{9/2}}{8 \1 p^{11/2}} + \frac{3 \1 \pi (449064+364712 e^2+21859 e^4) \1 a \3 M^{11/2} }{8 \1 p^{13/2} }  \Biggr) \nonumber \\[5pt]
  &\quad + \frac{3 \1 \pi \1 a^2}{p^2} + \frac{3 \1 \pi  \1 (25- 3 \1 e^2) \1 a^2 \1 M}{p^3} + \frac{15 \1 \pi \1  (492-2 \1 e^2-e^4) \1 a^2 \1 M^2}{8 \1 p^4} \nonumber \\[5pt]
  &\quad +\frac{105 \1 \pi (342+ 67 e^2-3 \1 e^4) \1 a^2 M^3}{4 \1 p^5} +\frac{315 \1 \pi (15768+7272 e^2 -123 e^4-2 e^6) \1 a^2 M^4}{64 \1 p^6}  \nonumber \\[5pt]
  &\quad +\frac{2079 \1 \pi (19224+15120 \1 e^2+365 \1 e^4 - 15 \1 e^6) \1 a^2 M^5}{64 \1 p^7} \nonumber \\[5pt]
  &\quad \mp \Biggl( \frac{12 \1 \pi \1 (3-e^2) \1 a^3 \1 M^{1/2}}{p^{7/2}} + \frac{20 \1 \pi (43 - 6 \1 e^2) \1 a^3 M^{3/2}}{p^{9/2}} +  \frac{21 \1 \pi (590+61 e^2 -9 \1 e^4) \1 a^3 M^{5/2}}{p^{11/2}} \nonumber \\[5pt]
  &\quad + \frac{9 \1 \pi (31500+12682 \1 e^2-897 e^4) \1 a^3 M^{7/2}}{2 \1 p^{13/2}} \Biggr) \nonumber \\[5pt]
  &\quad +\frac{9 \1 \pi  (3 - 2 \1 e^2) \1 a^4}{4 \1 p^4} + \frac{45 \1 \pi (41-12 \1 e^2+e^4)\1 a^4 M}{4 \1 p^5}  \nonumber \\[5pt]
  &\quad +\frac{35 \1 \pi (4966+15 \1 e^2-48 \1 e^4+e^6) \1 a^4 M^2}{16 \1 p^6} +\frac{315 \1 \pi (8742+2977 \1 e^2 -413 \1 e^4 +9 \1 e^6 ) \1 a^4 M^3}{16 \1 p^7} \nonumber \\[5pt]
  &\quad \mp \Biggl( \frac{15 \1 \pi (9 -4 \1e^2+e^4) \1 a^5 M^{1/2}}{p^{11/2}} +\frac{105 \1 \pi (59 - 6 \1 e^2 + e^4)\1 a^5 M^{3/2} }{p^{13/2}}  \Biggr) \nonumber \\[5pt]
  &\quad +\frac{3 \1 \pi (45-26 \1 e^2 +15\1  e^4 )\1 a^6}{8 \1 p^6} +\frac{21 \1 \pi (855-181\1 e^2+55 \1 e^4-5 \1 e^6) \1 a^6 M}{8 \1 p^7} + \cdots . \label{Kerr_PN_higher}
\end{align}
Note that the upper and lower signs denote the prograde and retrograde orbit, respectively. We can see that the effect of the BH spin first appears in the 3/2 order term.

\subsection{Chazy-Curzon spacetime}
Expanding Eq.~\eqref{cc_kin_f} in powers of $M/p$, we obtain the PN expansion formula for the periapsis shift in the Chazy-Curzon spacetime~\cite{Bini:2005dy,Vogt:2008zs}:
\begin{align}
  \Delta \phi_\text{C,PN} &= \frac{6 \pi M}{p} + \frac{\pi \1 (44 - 9 \1 e^2) M^2}{2 \1 p^2} +  \frac{\pi \1 ( 96 - 53 \1 e^2/2 ) M^3}{p^3} + \frac{\pi \1 (14064 - 4096 \1 e^2 +227 \1 e^4  ) M^4}{32 \1 p^4} \nonumber \\[5pt]
  &\quad + \frac{\pi \1 (11651 \1 e^4 - 250320 \1 e^2 + 992880) M^5}{480 \1 p^5}  \nonumber \\[5pt]
  &\quad + \frac{\pi \1 (11427648 - 1785648 \1 e^2 - 317876 \1 e^4 + 17529 \1 e^6) M^6}{1152 \1 p^6} \nonumber \\[5pt]
  &\quad + \pi \left( 48202  -\frac{1687 \1 e^2}{8} - \frac{1305073 \1 e^4}{240} + \frac{22478051 \1 e^6}{40320}   \right) \frac{M^7}{p^7} \nonumber \\[5pt]
  &\quad + \pi \left( \frac{11352407}{48} + 48083 \1 e^2 -\frac{1017922913 \1 e^4}{17280} + \frac{521182283 \1 e^6}{60480} - \frac{7450779 \1 e^8}{40960}    \right) \frac{M^8}{p^8} \nonumber \\[5pt]
  &\quad + \cdots . \label{cc_kin_pn_higher}
\end{align}
The higher order terms of the third order or more, which have not been given in Refs.~\cite{Bini:2005dy,Vogt:2008zs}, are presented here. Note that the orbital parameters $e$ and $p$ in Eq.~\eqref{cc_kin_pn_higher} are defined in the Weyl coordinate and these definitions differ from those in the Schwarzschild coordinate. The PN expansion formula in the Schwarzschild spacetime~\eqref{eq:sch_PN_higher} is rewritten in the Weyl coordinate as follows~\cite{Bini:2005dy,Vogt:2008zs}:
\begin{align}
  \Delta \phi_\text{S,PN} = \frac{6 \pi M}{p} + \frac{3( 14 - 3 \1 e^2) \pi M^2}{2 \1 p^2} + \cdots . \label{sch_kin_pn_higher_ver2}
\end{align}
Comparing Eqs.~\eqref{cc_kin_pn_higher} and \eqref{sch_kin_pn_higher_ver2}, we can see that the first order of $M/p$ is the same amount, while the difference occurs in the second order term. This difference is due to the multipole moments of the Chazy-Curzon spacetime.

\renewcommand{\theequation}{C.\arabic{equation}}
\makeatletter
\@addtoreset{equation}{section}
\makeatother
\section{Conditions for stable circular orbits in the Chazy-Curzon spacetime} \label{ap:CC_circular_condition}
Here, let us discuss the conditions for circular orbits of a timelike test particle and their stability in the Chazy-Curzon spacetime. 
We rewrite Eq.~\eqref{CC_rhodot} as
\begin{align}
  \dot{\rho}^2 = R(\rho) , \label{CC_rhodot_ver2}
\end{align}
where we have defined another effective potential as
\begin{align}
R(\rho) := \mathrm{e}^{M^2/\rho^2} \left[  \tilde{E}^2 - V(\rho) \right] . \label{CC_effective_pottential_R}
\end{align}
Differentiating both sides of Eq.~\eqref{CC_rhodot_ver2} with respect to the proper time $\tau$, we get
\begin{align}
  \ddot{\rho} = \frac{1}{2} R^\prime (\rho) ,
  \label{CC_rho_ddot_ver2}
\end{align}
where the prime denotes the derivative with respect to $\rho$. Note that
\begin{align}
  R^\prime (\rho) = -\frac{2 M^2 \mathrm{e}^{M^2/\rho^2}}{\rho^3} \left[ \tilde{E}^2-V(\rho ) \right] -\mathrm{e}^{M^2/\rho^2} V'(\rho ) .
\end{align}
From the above, the conditions for circular orbits $\dot{\rho}=0$ and $\ddot{\rho}=0$ yield
\begin{align}
  \tilde{E}^2 = V(\rho), \quad V^\prime (\rho) = 0,
\end{align}
respectively.

If the circular orbit at $\rho = \rho_0$ is perturbed as $\rho = \rho_0 + \varepsilon$ with $\varepsilon$ being infinitesimally small, we obtain
\begin{align}
  \ddot{\varepsilon} = \frac{1}{2} R^{\prime \prime} (\rho_0) \1 \varepsilon = - \frac{1}{2} \mathrm{e}^{M^2/\rho^2} V^{\prime \prime} (\rho_0 ) \1 \varepsilon ,
\end{align}
from Eq.~\eqref{CC_rho_ddot_ver2}. The condition for the circular orbit to be stable is that $\varepsilon$ obeys a simple harmonic motion. Then we can find the condition is $V^{\prime \prime} (\rho) \geq 0$.

\bibliography{refs}
\bibliographystyle{JHEP}
\end{document}